\documentclass[pra,twocolumn,preprintnumbers,amsmath,amssymb,superscriptaddress]{revtex4}
\usepackage{dsfont}
\usepackage{color}
\usepackage{graphicx}% Incl fgfrude figure files
\usepackage{dcolumn}% Align table columns on decimal point
\usepackage{bm}% bold math
\usepackage{hyperref}
\usepackage{enumitem}
\usepackage{textcomp}
\usepackage{balance}
\usepackage{comment}
\usepackage{cleveref}
\usepackage{bbold}
\usepackage{amsthm}

\begin{document}

\title{Macroscopic maximally entangled state preparation between two atomic ensembles}

\author{Manish Chaudhary}
\thanks{manish.manish11167@gmail.com}
\thanks{manish.phys123@gmail.com}
\affiliation{State Key Laboratory of Precision Spectroscopy, School of Physical and Material Sciences, East China Normal University, Shanghai 200062, China}
\affiliation{New York University Shanghai, 567 West Yangsi Road, Pudong, Shanghai 200126, China}

\author{Ebubechukwu O. Ilo-Okeke}
\affiliation{New York University Shanghai, 567 West Yangsi Road,
Pudong, Shanghai 200126, China}
\affiliation{Department of Physics, School of Physical Sciences,
Federal University of Technology, P. M. B. 1526, Owerri, Imo State 460001, Nigeria}

\author{Valentin Ivannikov}
\affiliation{New York University Shanghai, 567 West Yangsi Road,
Pudong, Shanghai 200126, China}
\affiliation{NYU-ECNU Institute of Physics at NYU Shanghai, 3663 Zhongshan Road North, Shanghai 200062, China}

\author{Tim Byrnes} 
\thanks{tim.byrnes@nyu.edu}
\affiliation{New York University Shanghai, 567 West Yangsi Road,
Pudong, Shanghai 200126, China}
\affiliation{State Key Laboratory of Precision Spectroscopy, School of Physical and Material Sciences, East China Normal University, Shanghai 200062, China}
\affiliation{NYU-ECNU Institute of Physics at NYU Shanghai, 3663 Zhongshan Road North, Shanghai 200062, China}
\affiliation{Center for Quantum and Topological Systems (CQTS),
NYUAD Research Institute, New York University Abu Dhabi, UAE}
\affiliation{Shanghai Frontiers Science Center of Artificial Intelligence and Deep Learning, 567 West Yangsi Road, Pudong New District, Shanghai 200126, China}
\affiliation{Department of Physics, New York University, New York, NY, 10003, USA}

\date{\today}% It is always \today, today,

\begin{abstract}
We develop a scheme to prepare a macroscopic maximally entangled state (MMES) between two atomic ensembles using adaptive quantum nondemolition (QND) measurements. %In adaptive QND scheme, we have analyzed how sequential QND measurements in conjunction with unitary correction operations ensure the convergence towards the maximally entangled macroscopic state. 
The quantum state of the system is evolved using a sequence of QND measurements followed by adaptive unitaries, such that the desired measurement outcome is obtained with asymptotically unit probability. This procedure is repeated in $z$ and $x$ spin basis alternately such that the state converges deterministically towards the maximally entangled state. 
%The ultimate effect of multiple QND measurements made in the alternative $z$ and $x$ basis is to derive the system towards a macroscopic maximally entangled state. 
Up to a local spin-basis rotation, the maximally entangled state has zero total spin angular momentum, i.e. it is a singlet state.
%and is not a random mixture of non entangled thermal states. This state exhibit larger correlations and have been proposed for a variety of applications in quantum information and computation. Our protocol works beyond the Holstein-Primakoff approximation as is conventionally done, and treats the spins of the atomic gases in an exact way. In addition, our scheme does not require to perform postselection on the measurements outcomes. 
Our protocol does not perform postselection and works beyond the Holstein-Primakoff regime for the atomic spin degrees of freedom, producing genuine macroscopic entanglement.  
\end{abstract}

%\pacs{03.75.Dg, 37.25.+k, 03.75.Mn}% PACS, the Physics and Astronomy
\maketitle

\section{\label{sec1}Introduction}
%Entanglement and one atomic ensemble entanglement review
Entanglement plays an important role in various quantum information tasks such as teleportation \cite{bennett1993teleporting}, cryptography \cite{quantumcryptography} and its production is one of the essential capabilities when constructing a quantum computer \cite{ladd2010quantum,mermin2007quantum,preskill2012quantum}. Entanglement is considered a resource in the context of quantum information science \cite{horodecki2009quantum,chitambar2019quantum,wilde2013quantum,bouwmeester2000physics,schleich2016quantum}. In the standard model of quantum computing, composite systems of qubits can be used to form a quantum register \cite{Nielsen:2010,mermin2007quantum}. However, quantum protocols based on higher dimensional systems have recently attracted a great attention \cite{chi2022programmable,bouchard2017high,briegelraussendorf,paolomaximally,PhysRevA.106.042614} and offer certain advantages such as a higher information capacity and increased resistance to noise \cite{wang2020qudits,PhysRevLett.88.127902,scottmultipartite,campbellfaulttolerant}. Higher-dimensional systems are advantageous as these allow for lower detection efficiency than qubits \cite{PhysRevLett.104.060401,quantumsteeringqudits}. Several physical systems allow for the encoding of higher-dimensional quantum information. These systems include Rydberg atoms \cite{omran2019generation}, trapped ions \cite{PhysRevX.5.021026}, cold atomic ensembles \cite{ding2016high}, superconducting phase qudits \cite{neeley2009emulation},  photonic systems \cite{kues2017chip,zhang2019quantum}, and mechanical resonators \cite{kotler2021direct}. 
Atomic gases are a particularly fascinating physical platform for observing many-body entanglement, due to the high level of controllability and low decoherence \cite{hammerer2010quantum,lukin2000entanglement}.  One of the most elementary type of entangled states for an atomic gas are spin squeezed states, where particular observables are reduced below the standard quantum limit \cite{sorensen2001many,hald1999spin,kuzmich2000generation}, and has numerous applications in quantum metrology \cite{gross2012spin,Giovannetti,giovannetti2004quantum,toth2014quantum,giovannetti2011advances,you2017multiparameter,bao2020spin,Sekatski2017quantummetrology}. It has also been observed that Bell violations \cite{bell1964,freedman1972experimental,aspect1982experimental}, which are a stronger form of quantum correlations in the quantum quantifier hierarchy \cite{adesso2016measures,ma2019operational}, can be generated in Bose-Einstein condensates (BECs) \cite{schmied2016bell,kitzinger2021bell}.

Maximally entangled states such as Bell states in a two qubit system \cite{einstein1935can,bell1964,mermin1993,Nielsen:2010} are of great importance for numerous quantum information tasks. Quantum communication schemes such as teleportation, dense coding, and entanglement swapping require control over a basis of maximally entangled quantum states \cite{bennett1993teleporting,densecoding1996,entanglementswapping1993}. In optical systems these states are routinely generated and detected \cite{o2009photonic}. In higher dimensions, maximally entangled states can potentially be used for the teleportation of more complex quantum states in the larger Hilbert space  \cite{sych2009complete,nparticlesinglet,luo2019quantum,hu2020experimental}.  While most of the work relating to entanglement in atomic ensembles has been focused on entanglement that exists between atoms in a single ensemble \cite{gross2012spin}, works extending this to two or more spatially separate ensembles have also been investigated both theoretically and experimentally {\cite{hammerer2010quantum,byrnes2013fractality,jing2019split,fadel2022multiparameter,vitagliano2023number}.
The first experimental demonstration of entanglement between atomic gases was observed in paraffin-coated hot gas cells \cite{julsgaard2001experimental} using quantum nondemolition (QND) measurements where the entanglement between two atomic ensembles had been produced in the form of two-mode squeezed states. For BECs, entanglement has been observed between two spatial regions of a single BEC \cite{esteve2008squeezing,kunkel2018spatially,fadel2018spatial}. %but never between two completely separate BECs, to date.  
Recently, entanglement between two separate BECs was also reported \cite{PhysRevX.13.021031}.
Such entanglement is fundamental to performing various quantum information tasks based on atomic ensembles, such as quantum teleportation \cite{krauter2013deterministic,pyrkov2014quantum,pyrkov2014full,braunstein2005quantum}, remote state preparation \cite{manish2021remote}, clock synchronization \cite{ilo2018remote}, and quantum computing \cite{byrnes2012macroscopic,byrnes2015macroscopic,abdelrahman2014coherent}.  
In the past, numerous theoretical and experimental works has been focused on generating macroscopic singlet states within single atomic ensembles using collective QND measurement \cite{behbood2014,toth2010generation,behbood2013feedback}. 
%The generation of MMES for many-body systems is a fascinating research field as these states appear as the ground states of many fundamental spin models \cite{anderson1987resonating,}. 
This state is basis invariant that finds considerable importance in quantum information processing \cite{bennett1993teleporting,entanglementswapping1993,densecoding1996,balents2010spin,PhysRevLett.105.013603}. Currently, the amount of entanglement that can be experimentally generated is very small, working within the Holstein-Primakoff approximation of spins, such that Hilbert space of the spins is largely unused. As such, current experiments are far below levels where a MMES can be generated even in principle from the way the protocols are constructed.   
%Out of the sets for maximally entangled state in spin systems, there are states for which the collective spin angular momentum is zero, known as singlet states. 

%In the past, numerous theoretical and experimental works has been focused on generating macroscopic singlet state (MSS) within single atomic ensembles using collective quantum nondemolition (QND) measurement \cite{behbood2014,toth2010generation,behbood2013feedback}. In a QND scheme, the atoms in ensemble interact with a light field, which is subsequently measured projecting the atoms into a entangled state \cite{julsgaard2001experimental,kuzmich2000generation}. Out of the sets for maximally entangled state in spin systems, those states for which the collective spin angular momentum is zero, are interesting as these macroscopic maximally entangled states appear as ground states of many fundamental spin models \cite{anderson1987resonating,balents2010spin}. Moreover the singlet state is basis invariant that finds considerable importance in quantum information processing \cite{bennett1993teleporting,entanglementswapping1993,densecoding1996}. 

%However, there is no experimental evidence of generation of MMES for two separated atomic clouds so far.
In this paper we propose a scheme for the generation of a MMES between two atomic ensembles using collective QND measurement and local spin rotations. In the QND scheme, the atoms in ensemble interact with a photonic field, which is subsequently measured, projecting the atoms into an entangled state \cite{aristizabal2021quantum,manish2022measurement}. Our approach extends works which have proposed sequential QND measurements to generate a collective singlet state within single atomic ensembles with postselection methods such as in Ref. \cite{behbood2014,behbood2013feedback} and using feedback techniques \cite{toth2010generation}. Our scheme, on the other hand, is deterministic in the sense that the system converges towards MMES with \textit{asymptotically unit} probability as opposed to the stochastic evolution based on the random measurement outcomes in the sequential QND \cite{manish2022measurement}. 
%\textcolor{red}{Feedback is avoided in most spin-squeezing experiments, since more often conditional spin expectation values are measured. This makes these experiments simpler. Our scheme merely uses adaptive strategy using spin rotations}. 
Our scheme does not approximate spins as a bosonic mode under the Holstein-Primakoff approximation as is often done by restricting to the short time interaction regime and holds for longer evolution times. In addition, our scheme does not rely upon individual atom control, as we have employed collective spin operations, projective measurements and local unitary rotations that can be implemented in experimental settings.

The paper is structured as follows.  In Sec. \ref{sec2} we review the theory of QND measurement induced entanglement \cite{aristizabal2021quantum,manish2022measurement} and introduce the basic physical system that we are dealing with. In Sec. \ref{sec3} we describe the maximally entangled state for macroscopic atomic ensembles and show its connection to the macroscopic singlet state. The former can be transformed into the latter state through a local unitary transformation. In Sec. \ref{sec4} we explain the protocol for deterministic preparation of the MMES and show that multiple sequential QND measurement produces a convergence of the desired state with the adaptive unitary.  In Sec. \ref{sec5} we numerically simulate our proposed protocol and show that convergence is obtained towards the MMES. In Sec. \ref{sec6}, we have discussed the overall effectiveness of the protocol with imperfections such as atom number fluctuations and initial ensemble prepared in the maximally mixed state. In Sec. \ref{sec7}, we propose an experimental set-up to realize the protocol. Finally, in Sec. \ref{sec8} we summarize our results.

\section{QND measurements}
\label{sec2}
Here we review the theory of QND measurements on the atomic ensembles as introduced in Ref. \cite{aristizabal2021quantum}. The effect of multiple such QND entanglement operations is studied in Ref. \cite{manish2022measurement}.

\subsection{Definitions and Physical system}
 The physical system we shall consider consists of two neutral atomic ensembles or BECs, where each atom has two populated internal states.  A common choice for the internal states are hyperfine ground states,
 such as the $ F= 1, m_F = - 1 $ and $ F = 2, m_F = 1 $ states in the case of $^{87}$Rb
\cite{pezze2018quantum}. For BECs we denote the bosonic annihilation operator for the two states as $ g_l, e_l $ respectively, where $ l \in \{1,2 \} $ labels the two BECs.  These operators can be used to define an effective spin using the Schwinger boson operators 
\begin{align}\label{eq:spinops}
    S^x_l & =e_l^\dagger g_l+g_l^\dagger e_l \nonumber \\
    S^y_l & =-ie_l^\dagger g_l+ig_l^\dagger e_l \nonumber \\
     S^z_l & =e_l^\dagger e_l-g_l^\dagger g_l  . 
\end{align}
The commutation relation for the spin operators are
\begin{align}
[S^j,S^k]=2i\epsilon_{jkl}S^l,   \label{commutator}
\end{align}
where $\epsilon_{jkl}$ is the Levi-Civita symbol.

For atomic ensembles, the total spin operators are written in terms of collective spin operators 
\begin{align}\label{eq:spinopsensemble}
    S^x_l & =  \sum_{n=1}^N \sigma^x_{l,n}  \nonumber \\
    S^y_l & =\sum_{n=1}^N \sigma^y_{l,n}  \nonumber \\
     S^z_l & =\sum_{n=1}^N \sigma^z_{l,n},
\end{align}
where $ \sigma^k_{l,n} $ is a Pauli operator for the $n$th atom in the $l$th ensemble. 
For simplicity, we consider that the number of atoms $N$ in each ensemble are equal. For the case that all the operations on the atomic ensembles are completely symmetric under particle interchange from the initialization of the states to the final measurement, the formalism (\ref{eq:spinops}) and (\ref{eq:spinopsensemble}) for the BECs and atomic ensembles respectively are completely equivalent \cite{timquantumoptics2020}.  We will use the bosonic formulation (\ref{eq:spinops}) henceforth, although it should be understood that our calculations apply to both the BEC and atomic ensemble case. \newline

The spin coherent states for $N$ uncorrelated atoms in an ensemble is defined as
\begin{align}
    | \theta , \phi \rangle\rangle_l =\frac{( \cos \frac{\theta}{2} e^{-i \phi /2 }  e_l^\dagger+ \sin \frac{\theta}{2} e^{i \phi /2 } g_l^\dagger)^N}{\sqrt{N!}}| \text{vac} \rangle
    \label{spincoherent}
\end{align}
where $\theta , \phi$ are the angles on the Bloch sphere, and $ | \text{vac} \rangle $ is the vacuum state containing no atoms. The Fock states are defined as 
\begin{align}
|k \rangle_l = \frac{(e_l^\dagger)^k (g_l^\dagger)^{N-k}}{\sqrt{k! (N-k)!}} | \text{vac} \rangle .
\label{fockstates}
\end{align}
The Fock states are eigenstates of the $ S^z $ operator according to
\begin{align}
S^z_l |k \rangle_l =  (2 k-N) |k \rangle_l  .
\end{align}

\subsection{QND Entanglement}
Here we summarize the elementary entangling operation that we will use in our protocol for deterministic preparation of maximally entangled states. Coherent light is used to perform an indirect measurement of two atomic ensembles arranged in a Mach-Zehnder configuration (Fig. \ref{fig1}). The atoms in the ensemble are prepared in a product state of two spin coherent states and the interaction between photons and atoms is governed by the Hamiltonian \cite{kuzmich2000generation}, 
\begin{align}
	H = \kappa(S^z_{1}-S^z_{2})J^z ,
	\label{interhamiltonian}
\end{align}
where $\kappa$ is the coupling constant and $J^z = a_1^\dagger a_1 - a_2^\dagger a_2 $ is the Stokes operator for the two optical modes $a_1,a_2$ that enter into each arm of the interferometer. 

After interacting with the atoms, the photonic modes are interfered with a beam splitter, giving rise to new modes $c,d$ and the photons are detected by the detectors with counts $n_c,n_d$ respectively. After the measurement, the atomic ensembles collapse in the $S^z_1-S^z_2$ spin observable basis \cite{aristizabal2021quantum,manish2022measurement}.
%
%Both the BECs are prepared initially in $S^x$-polarized states (\ref{spincoherent})
%\begin{align}
   % |\psi_0\rangle & = \Big|\frac{\pi}{2},0\Big\rangle \Big\rangle_1 \Big|\frac{\pi}{2},0\Big\rangle \Big\rangle_2 \nonumber\\
  %  & = \frac{1}{2^N}\sum_{k_1,k_2} \sqrt{{N \choose k_1} {N \choose k_2}   } |k_1,k_2\rangle
  %  \label{initialstate}
%\end{align}
%
%

As shown in Ref. \cite{manish2022measurement}, the QND entanglement scheme between two atomic ensembles can be described in terms of a Positive Operator Valued Measure (POVM) as
\begin{align}
M_{n_c n_d}(\tau) =\sum_{k_1,k_2 = 0}	C_{n_c,n_d}[(k_1-k_2)\tau]|k_1,  k_2\rangle \langle k_1 , k_2| ,
\label{povmdef}
\end{align}
where the modulating function is defined as
\begin{align}
	C_{n_c,n_d}(\chi) = \frac{\alpha^{n_c+n_d}e^{-|\alpha|^2/2}}{\sqrt{n_c!n_d!}}\cos^{n_c}(\chi)\sin^{n_d}(\chi),
 \label{cfunc}
\end{align}
and $\tau = \kappa t$ is the interaction time. The resulting state after the measurement is
\begin{align}
|\widetilde{\psi}_{n_c n_d} (\tau) \rangle & = M_{n_c n_d}(\tau) | \psi_0 \rangle  \nonumber \\
& = \sum_{k_1, k_2} \langle k_1, k_2 | \psi_0\rangle C_{n_c, n_d} [ (k_1 - k_2) \tau] | k_1, k_2 \rangle .
\label{psichange}
\end{align}
According to the Eq. (\ref{psichange}), the initial wave function is modulated by an extra factor of $C_{n_c,n_d}[(k_1-k_2)\tau]$ which can result in a measurement-induced generation of entanglement.
\begin{figure}[t]%
	\includegraphics[width=\linewidth]{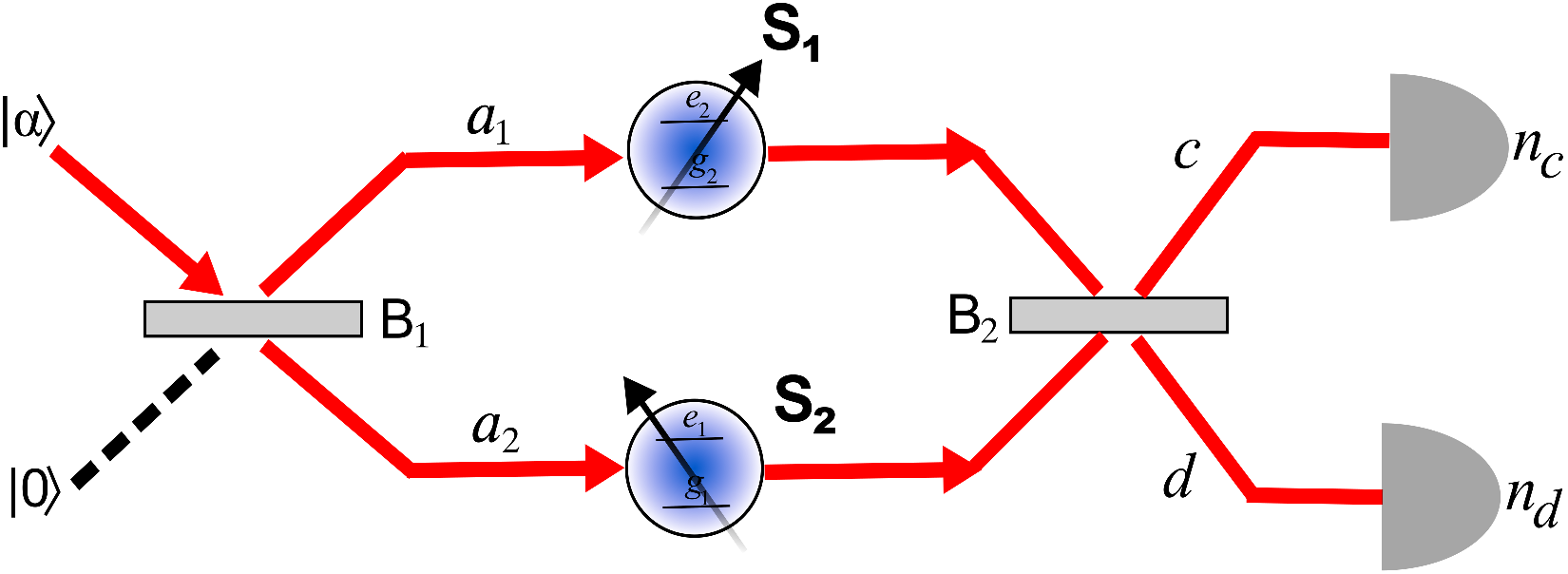}
	\caption{Entanglement generation between two atomic ensembles using the QND scheme. Coherent light $|\alpha\rangle$ is used to interact two-mode BECs via the QND Hamiltonian interaction (\ref{interhamiltonian}) arranged in a Mach-Zehnder configuration. The photon mode detections $n_c,n_d$ after the second beam splitter $B_2$ entangles the two spins $\textbf{S}_1$ and $\textbf{S}_2$.	}
	\label{fig1}%
\end{figure}

For large photon counts, the modulating function $C_{n_c, n_d} [(k_1 - k_2) \tau ] $ takes a Gaussian form \cite{aristizabal2021quantum} and is sharply peaked at
%The various photonic measurement outcomes are related to the states for BECs as,
%
\begin{align}
\sin^2[(k_1-k_2)\tau] = \frac{n_d}{n_c+n_d}.
\label{deltancnd}
\end{align}
%

%Being  $  M_{n_c n_d}^\dagger (\tau)  M_{n_c n_d} (\tau) $ to be a positive operator, it satisfies the definition of being a POVM. 

Taking the interaction time $ \tau  = \pi/2N $ and assuming $| \alpha \tau |^2 \gg 1 $,  as defined in \cite{manish2022measurement}, we may then approximate the POVM (\ref{povmdef}) as a measurement operator according to
\begin{align}
M_{n_c n_d} ( \tau= \frac{\pi}{2N}  ) \approx  
\Pi_\Delta ,
\end{align}
where the projections $ \Delta = k_1-k_2 $ and photonic measurements $ n_c, n_d $ are related according to (\ref{deltancnd}), and we defined
\begin{align}
\Pi_\Delta =&\frac{1}{2^{ \delta_{\Delta} }} \Big( \sum_{k=0}^{N-\Delta}|k,k + \Delta\rangle \langle k,k+\Delta|  \nonumber\\
& + (-1)^{(1- \delta_{\Delta})n_d} \sum_{k'=\Delta}^N |k',k'-\Delta\rangle \langle k',k' -\Delta|  \Big) . 
\label{projzbasis}
\end{align}
Here $ \delta_{\Delta} $ is the Kronecker delta which is 1 if $ \Delta = 0 $ and 0 otherwise. 

As it is clear from the definition of modulating function (\ref{cfunc}) and noted in Ref. \cite{manish2022measurement}, there is a sign difference between the two terms for odd $n_d$ photonic measurements. 
%Hence, there are two measurement operator defined for outcomes $\Delta > 0$.
%\begin{align}
 %   \Pi_\Delta = \left\{
%\begin{array}{cc}
%&\Pi^{+}_\Delta, \hspace{0.5cm} n_d=\text{even}\\
%&\Pi^{-}_\Delta, \hspace{0.5cm} n_d=\text{odd}
%\end{array}
%\right.  
%\label{plusminusproj}
%\end{align}
Since the shot-to-shot photonic outcome $n_d$ is random, the two measurements (\ref{projzbasis}) occur randomly and leads to stochastic evolution of the system. An exception is the outcome $ \Delta =0 $ which is independent of photonic count $ n_d $. We will show that in our protocol it is possible to construct an adaptive unitary that is independent of $n_d$ (and thus avoids explicit photon counting) and still converges towards the MMES. 

The measurement operators are defined in different spin bases by applying suitable unitary rotation as \cite{timquantumoptics2020}
\begin{align}
\Pi_\Delta^{(\theta,\phi)} =  {\cal U}  (\theta, \phi) \Pi_\Delta^{(z)}  {\cal U}^\dagger (\theta, \phi)
\label{rotatedproj}
\end{align}
where 
\begin{align}
{\cal U} (\theta, \phi) =  e^{-i (S^z_1+S^z_2) \phi /2}  e^{-i (S^y_1 + S^y_2) \theta/2} ,
\label{unitaryrotation}
\end{align}
and $ \Pi_\Delta^{(z)} $ is the same measurement operator as in (\ref{projzbasis}), but we explicitly specified the basis with the $ ^{(z)} $ label. \newline

\section{The Maximally entangled state}
\label{sec3}
In this section we discuss the nature of the maximally entangled state between two BECs. Namely, we would like to create the state,
\begin{align}
|\text{MMES} \rangle = \frac{1}{\sqrt{N+1}} \sum_k |k\rangle_1 |k \rangle_2. 
\label{eprstate}
\end{align}
This state has an entanglement of $ E = \log_2 (N+1)$ using the von Neumann entropy, which is the maximum value for two $N+1$ level systems. This state is also known as the spin-EPR state for atomic ensembles \cite{kitzinger2020}. 

We now show that the MMES (\ref{eprstate}) has a very close connection with the spin zero singlet state.  This fact shall be used to construct our protocol.  Each BEC can be considered to be a macroscopic qubit state with spin value $s_1 = s_2 = N/2$. Due to each boson being symmetric under interchange, the total spin is always in the maximum spin sector. For two spins, %there is a unique singlet state $|\mathcal{ S}_N \rangle$ 
one can define the collective state
that can be formed, with quantum numbers of the total spin $\bm{s}_{\text{tot}} = \bm{s}_1+\bm{s}_2$, here we have used the notation $ \bm{s}_l = \bm{S}_l/2 $, to connect our notation to the standard conventions of quantum angular momentum and $l\in \{1,2\}$ labels two atomic ensembles. 
%$\bm{S}_{\text{tot}}= \bm{S_1}+ \bm{S_2} $.  
%
%where $|S_N \rangle$ is the total spin eigenstates for two BECs each with atom number $N$.
We can explicitly write this state in terms of the total angular momentum eigenstate $| s, m\rangle $ where the two spins are coupled with  $ m = m_{1} + m_{2} $, $m$ is the orientation of total spin quantum number $s$ along $z$-direction such that,
\begin{align}
    & (s^z_1 +s^z_2 )  | s, m\rangle = m | s, m\rangle \nonumber \\ 
    & \bm{s}_{\text{tot}}^2 | s, m\rangle = s(s+1) | s, m\rangle.
\end{align}
There is a unique singlet state $| s_0, m_0\rangle$ which satisfies
\begin{align}
    & (s^z_1 +s^z_2 )  | s_0, m_0\rangle = 0\nonumber\\ 
    & \bm{s}_{\text{tot}}^2 | s_0, m_0 \rangle = 0 , 
    \label{singletcondition}
\end{align}
with $s_0=m_0=0$.
Using the coupling rule for two spins \cite{weissbluth2012atoms}, the singlet state then reads 
\begin{align}
 |s_0 , m_0 \rangle &=\sum_{m} \frac{(-1)^{s-m}}{\sqrt{2s+1}} |s, m\rangle_1 |s, -m \rangle_2 .
\label{singletstate}
\end{align}
%
 %For two qubits which are correlated, maximally entangled state with net spin momentum zero is defined as,
%
%\begin{align}
%|S_2\rangle =\frac{ |1\rangle_1|0\rangle_2 - |0\rangle_1|1\rangle_2}{\sqrt{2}}
%\label{singlet}
%\end{align}
%
%It is evident that by performing local transformations $U_{12} = $  \(  \mathbb{1} \)$\otimes e^{-i\sigma^y_2 \frac{\pi}{2}} $, the singlet state (\ref{singlet}) can be modified in to the state that is a possible outcome of QND measurements (\ref{deltavectoreven}),
%
%\begin{align}
%	\overline{|S_2\rangle} =\frac{ |1\rangle_1|1\rangle_2 + |0\rangle_1|0\rangle_2}{\sqrt{2}}
%	\label{singletstate}
%\end{align}
%
The state (\ref{singletstate}) has perfect correlations and anti-correlations in the linear combination of spin observables. The state could be realized as the ground spin state of the Hamiltonian $ \bm{S}^2 $.

For atomic ensembles of collection of $N$ atoms, we describe the state in Fock space (\ref{fockstates}) that can be equivalently described in the angular momentum basis as well
\begin{align}
|k \rangle = \Big|s=\frac{N}{2},m = k-\frac{N}{2} \Big\rangle,
\label{angmombasis}
\end{align}
\newline
%For two BECs one could write it as
%
%\begin{align}
%& |k_1,k_2 \rangle   = \nonumber \\ 
%& \Big |s_1=\frac{N_1}{2},m_1 = k_1-\frac{N_1}{2}; s_2=\frac{N_2}{2},m_2 = k_2-\frac{N_2}{2}\Big\rangle 
%\label{angmombasis}
%\end{align}
The singlet state (\ref{singletstate}) is defined for atomic ensemble using the relation (\ref{angmombasis})
\begin{align}
    |s_0 , m_0 \rangle & =  \frac{1}{\sqrt{N+1}}\sum_{k=0}^N (-1)^k |k\rangle_1 |N-k\rangle_2.
%& =  \sum_{m}  \frac{(-1)^{s-m}}{\sqrt{2s+1}} \Big |s=\frac{N}{2},m = k-\frac{N}{2}\Big\rangle_1 \nonumber\\ 
%& \Big|s=\frac{N}{2},m = -\Big(k-\frac{N}{2}\Big)\Big\rangle_2 \nonumber\\
\label{singletstatebec}
\end{align}
%
%The state (\ref{singletstatebec}) could be expressed in terms of (\ref{angmombasis}) as
%\begin{align}
 %  |s_1,m_1; s_2,m_2\rangle = \sum_{S}  |S, M \rangle \langle S, M|s_1,m_1; s_2,m_2\rangle
%\end{align}
%The states $ |S, M_S\rangle $ are the eigenstates of the total Hamiltonian $ H = (\bm{S}_1+\bm{S}_2)^2 $ which could be viewed as effective total spin, while the ground state of this spin Hamiltonian is the singlet state (\ref{singletstatebec}). 

We see that there is a close connection between the maximally entangled state (\ref{eprstate}) and the singlet state (\ref{singletstatebec}).  In fact, the singlet state is a MMES up to a local basis transformations. 
The local spin basis rotation, 
\begin{align}
    e^{-iS^y_2\frac{\pi}{2}} |k\rangle  = (-1)^k|N-k\rangle,
\end{align}
transforms the singlet state to the maximally entangled state as
\begin{align}
    |\text{MMES}\rangle & = e^{-iS^y_2\frac{\pi}{2}} |s_0 , m_0 \rangle . 
%& =\frac{1}{\sqrt{2s+1}} \sum_{m_s} \Big |s=\frac{N}{2},m_s = k-\frac{N}{2}\Big\rangle_1 \nonumber\\
%&\Big|s=\frac{N}{2},m_s = k-\frac{N}{2}\Big\rangle_2
\label{transformedsingletstatebec}
\end{align}
%
%that has the same form as MMES state (\ref{eprstate}). 
%

%The state (\ref{transformedsingletstatebec}) could be obtained by QND scheme that has potential application in many-body spin systems. It is the lowest energy eigenstates of interaction Hamiltonian that  has spin correlations exactly same as QND scheme  
From (\ref{transformedsingletstatebec}) we may deduce the operator that has the analogous relation as (\ref{singletcondition}) for the MMES.  Applying the operator $ e^{-i S^y_2 \pi/2 } $ to (\ref{singletcondition}) and using (\ref{transformedsingletstatebec}) we have
\begin{align}
 e^{-i S^y_2 \pi/2} \bm{s}_{\text{tot}}^2 e^{i S^y_2 \pi/2} | \text{MMES} \rangle = \bar{\bm{s}}_\text{tot}^2 | \text{MMES} \rangle = 0 
 \end{align}
%\begin{align}
%\overline{\bm{S}}_{\text{tot}}^2 | %\text{EPR} \rangle = 0 
%\end{align}
where
\begin{align}
\bar{\bm{s}}_\text{tot}^2 = \frac{(S^x_1-S^x_2)^2 + (S^y_1+S^y_2)^2 + (S^z_1-S^z_2)^2}{4}
\label{totalspintwobec}
\end{align}
%This can be equivalently looked up as the joint eigenstate of interaction Hamiltonian
%\begin{align}
% H_{\text{int}} =  (S^x_1-S^x_2)^2 + (S^y_1+S^y_2)^2 + (S^z_1-S^z_2)^2
%\end{align}
% 
has same correlations in the spin observables as seen in QND interactions \cite{aristizabal2021quantum,manish2022measurement}. 
%It is obtained from (\ref{totalspintwobec}) by rotating the total spin operators to $S^x_1 - S^x_2$ and $S^z_1 - S^z_2$.\newline

For a two qubit system, the maximally entangled state (\ref{eprstate}) is the Bell state
\begin{align}
   \frac{|0\rangle_1 |0\rangle_2 + |1\rangle_1 |1\rangle_2}{\sqrt{2}}.
    \label{eprstatequbit}
\end{align}
This state is an eigenstate of the operators $ \sigma^z_1 - \sigma^z_2 $ and $ \sigma^x_1 - \sigma^x_2 $ with zero eigenvalue. 
%which is transformed locally to singlet state 
%\begin{align}
  %  |\mathcal{S}_1\rangle = \frac{|0\rangle_1 |1\rangle_2 - |1\rangle_1 |0\rangle_2}{\sqrt{2}}
%\end{align}
%In the rest of paper, we are dealing with the eq. (\ref{transformedsingletstatebec}) for macroscopic maximally entangled state and discuss the approach to the deterministic generation of the state. Fig. \ref{fig1} shows the schematic of QND process for generating entanglement between two atomic ensembles. The spin state of a particular ensemble is described by the collective spin vector (\ref{eq:spinopsensemble}) and the correlation for maximally enatangled spin state is generated using adaptive QND.
%however there is difference between the state originating from the random mixture of total spins (Fig. \ref{fig1}(a)) and the singlet pairs (Fig. \ref{fig1}(b)). The net spin angular momentum in both cases equal to zero, however there are no correlations set up between the atoms in the former case. In our scheme we intend to prepare the latter state on a macroscopic scale using QND measurement induced entanglement.

\section{Deterministic preparation of maximally entangled state}
\label{sec4}
As discussed in Sec. \ref{sec2}, QND measurements can be used to entangle two different atomic ensembles or BECs. Depending on the photonic measurement outcomes, the state of BEC collapses on different entangled states in general (\ref{psichange}). For instance, an initial state $|\psi_0\rangle$ is collapsed by measurement (\ref{projzbasis}) as  
\begin{align}
\Pi_{\Delta}^{(z)} | \psi_0 \rangle = \sum_{k=0}^{N-\Delta} \psi_{k}^{+} |k + \Delta \rangle | k  \rangle  + 
\sum_{k'=\Delta}^{N} \psi_{k'}^{-} |k' - \Delta \rangle | k' \rangle  
\label{projinitial}
\end{align}
where the coefficients in (\ref{projinitial}),
\begin{align}
    \psi_{k}^{+} & = \frac{1}{2^{ \delta_{\Delta} }}\langle k + \Delta| \langle k  |\psi_0\rangle \nonumber \\
    \psi_{k}^{-} & = \frac{(-1)^{(1- \delta_{\Delta})n_d}}{2^{ \delta_{\Delta} }}\langle k - \Delta| \langle k |\psi_0\rangle,
    \label{coefftransformedstate}
\end{align}
which is an entangled state for a particular measurement outcome $\Delta$. It is however not a MMES due to the amplitudes $ \psi^\pm_k $  being not necessary of equal magnitude, and the difference $ \Delta $ between the Fock states in the BECs. Our aim now will be to devise a protocol such that the MMES (\ref{eprstate}) can be prepared deterministically, using quantum measurements which are inherently random.  

\subsection{Basic idea}
To gain some intuition about the protocol that we will introduce later, let us introduce some basic properties of the QND measurements and the MMES.

The MMES is a unique state that is an eigenstate of both the measurement operators $\Pi_0^{(z)}$ and $\Pi_0^{(x)}$, 
 \begin{align}
   &\Pi_0^{(z)} |\text{MMES}\rangle =  |\text{MMES}\rangle\nonumber\\
   &\Pi_0^{(x)} |\text{MMES}\rangle =  |\text{MMES}\rangle . 
   \label{singleqndoperator}
 \end{align}
It then follows that an alternating sequence of such measurements has the $|\text{MMES}\rangle$ as an eigenstate 
\begin{align}
(\Pi^{(x)}_0 \Pi^{(z)}_0 )^M |\text{MMES} \rangle = |\text{MMES}\rangle .
\end{align}
Due to the unique nature of the MMES satisfying (\ref{singleqndoperator}), the QND measurements (\ref{projzbasis}) applied alternately on an arbitrary state $|\psi_0\rangle $ converges to the MMES (\ref{eprstate}),
\begin{align}
(\Pi_0^{(x)}\Pi_0^{(z)} )^M | \psi_0 \rangle \xrightarrow{M \rightarrow \infty }  |\text{MMES}\rangle.
\label{convergenceepr}
\end{align}
According to (\ref{singleqndoperator}), since the MMES is an eigenstate of both $ \Pi^{(z)}_0 $ and  $ \Pi^{(x)}_0 $ measurement operators, once the state $ |\text{MMES}\rangle $ is obtained, further application of the measurement operators do not change the state.  This is in fact an unique state for the same reasons that a singlet state is a unique state for two $ s_l= N/2 $ spins. Therefore it is a fixed point of the evolution. The MMES is obtained for the QND measurement (\ref{projzbasis}) corresponding to outcome $\Delta = 0$. However,  Eq. (\ref{convergenceepr}) does not constitute a physically realizable protocol because obtaining the $\Delta=0$ measurement outcome is set by Born's probability rule and due to the randomness of quantum measurements, we cannot guarantee that only the $\Delta=0$ outcome will be obtained.  

In order to overcome the randomness of quantum measurements and make a deterministic scheme, we use an adaptive strategy.  Our scheme involves applying a unitary transformation to the state in the event that a $\Delta\ne 0$ is obtained, and repeating the measurements many times until the desired $\Delta=0$ outcome is obtained. The protocol is deterministic in the sense that eventually a measurement sequence will always end up with the $\Delta=0$ outcome. The adaptive unitary is chosen such as to maximize the probability of obtaining the $\Delta = 0$ outcome in the next step. Our approach can be considered a special case of the measurement-based imaginary time evolution protocol proposed in Ref. \cite{mao2022imaginary,kondappan2022imaginaryqnd}.

\subsection{Protocol}
Here we more concretely describe the full procedure for deterministic preparation of the MMES using sequential QND measurements performed in $z$ and $x$ basis.

\begin{figure}%
	\includegraphics[width=\linewidth]{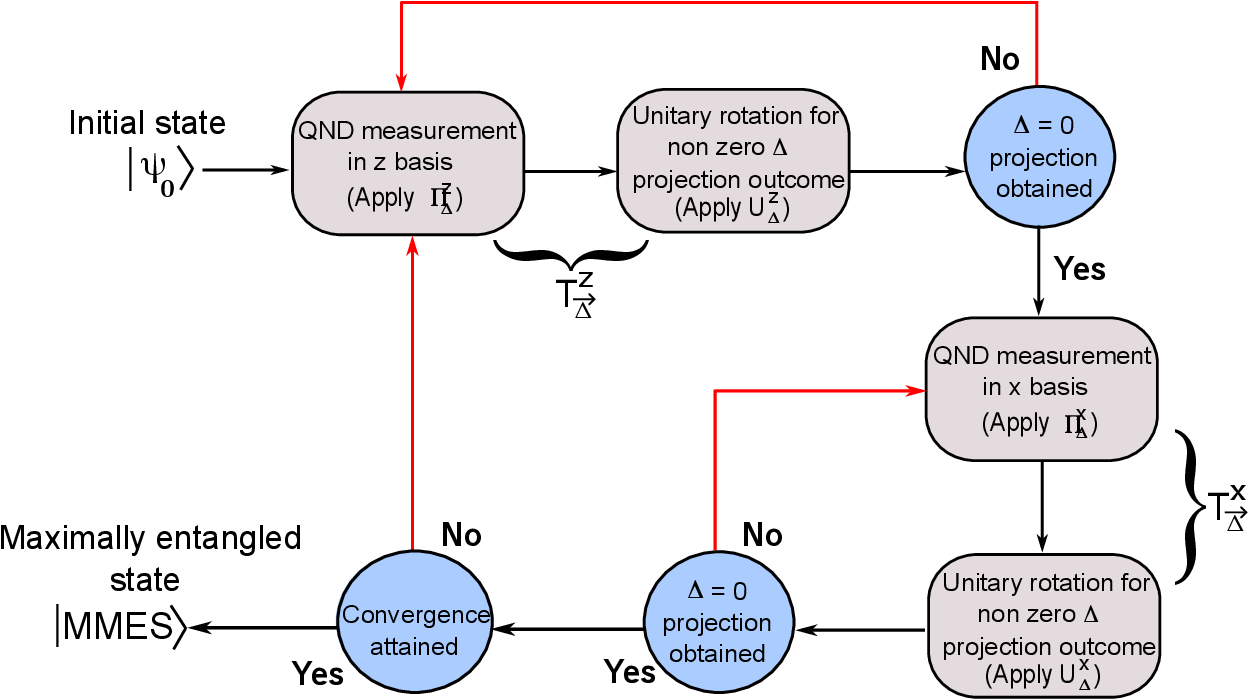}
	\caption{Protocol for obtaining the MMES. A ``repeat-until-success" measurement sequence $ T^{(z)}_{\vec{\Delta}} $ is applied to an initial state, where a sequence of projective measurements $ \Pi^{(z)}_\Delta $ and adaptive unitary rotations are made until the $ \Delta = 0 $ result is obtained.  The same repeat-until-success sequence is repeated in the $ x $ basis.  The two sequences are repeated until convergence is attained, where both $ z $ and $ x $ measurements yield $ \Delta = 0 $ on the first measurement.  This procedure converges to the MMES (\ref{eprstate}).  
 %An initial state is projected to an entangled state using QND measurement scheme, the projected state is measured and corrected alternately in $z$ and $x$ spin basis using sequential QND measurement outcomes (\ref{multipleconvergent})  that converges to a desired state (\ref{eprstate}).
	}
	\label{fig3}%
\end{figure}
We define the ``repeat-until-success" adaptive QND scheme which applies a sequence of QND measurements (\ref{projzbasis}) and unitary operators until the measurement outcome $\Delta=0$ is obtained as 
\begin{align}
    T_{\vec{\Delta}}^{(z)} & = \prod_{j=1}^{L} U_{\Delta_j}^{(z)} \Pi_{\Delta_j}^{(z)} \nonumber \\
& = \Pi_0^{(z)} U_{\Delta_{L-1}}^{(z)} 
\Pi_{\Delta_{L-1} }^{(z)} \dots 
U_{\Delta_{1}}^{(z)} 
\Pi_{\Delta_{1}}^{(z)},
\label{projsequence}
\end{align}
%
%such that 
%\begin{align}
  %  | \tilde{\psi}_{\vec{\Delta} } \rangle & = T_{\vec{\Delta}}^{(z)}|\psi_0\rangle. 
 %   \label{multizseq}
%\end{align}
where $ \Delta_L = 0 $ and $ U_0^{(z)} = I $.  
A particular repeat-until-success measurement sequence is labeled according to the notation,
\begin{align}
\vec{\Delta}=(\Delta_1,\Delta_2\dots\Delta_L).
\label{multizseq}
\end{align}

In order to make the state convergent towards MMES, we aim to correct those projections ($\Delta\neq0$) through unitary $U^{(z)}_{\Delta}$ that ensures the convergence,
\begin{align}
\Pi_0^{(z)} | \tilde{\psi}_{\vec{\Delta}} \rangle =  | \tilde{\psi}_{\vec{\Delta}} \rangle
\end{align}
where the unnormalized state after the repeat-until-success sequence is
\begin{align}
    | \tilde{\psi}_{\vec{\Delta} } \rangle & = T_{\vec{\Delta}}^{(z)}|\psi_0\rangle .
   \label{multizconvergence}
\end{align}
%
%\begin{align}
   % \prod_{j=1}^L (U^{(z)}_{\Delta^z_j}\Pi^{(z)}_{\Delta^z_j })|\psi_0\rangle \to \Pi^{(z)}_{0}|\psi_0\rangle . 
  %  \label{multizconvergence}
%\end{align}
%After applying another QND measurement in x basis and corresponding corrections yields the final state 
%
%
%\begin{align}
   % | \tilde{\psi}_{\vec{\vec{\Delta}} } \rangle & = T_{\vec{\Delta}}^{(x)}T_{\vec{\Delta}}^{(z)}|\psi_0\rangle. 
  % \label{multiplexzseq}
%\end{align}
%
%\begin{align}
  %  \Big (\prod_{k=1}^L U^{(x)}_{\Delta^x_k}\Pi^{(x)}_{\Delta^x_k }\Big)\Big(\prod_{j=1}^L U^{(z)}_{\Delta^z_j}\Pi^{(z)}_{\Delta^z_j }\Big)|\psi_0\rangle \to \Pi^{(x)}_{0}\Pi^{(z)}_{0}|\psi_0\rangle .  
  %  \label{multiplexzseq}
%\end{align}
%
%It is characterized by projection outcomes,
%\begin{align}
  %  \vec{\vec{\Delta}} = (\vec{\Delta}^z,\vec{\Delta}^x)
%\end{align}
%where the single arrow over $\Delta$ projection outcome follows from (\ref{multizseq}) and shows the various projections to achieve convergence in one basis.

Then analogously to (\ref{convergenceepr}), we replace each of the projectors in the $z$ and $x$ basis with the measurement sequences (\ref{projsequence}) such that
\begin{align}
    | \tilde{\psi}_{\vec{\vec{\Delta}} }^f \rangle & = \prod_{r=1}^{M}(T_{\vec{\Delta}_r^x}^{(x)}T_{\vec{\Delta}_r^z}^{(z)})|\psi_0\rangle \xrightarrow{M \rightarrow \infty }
    |\text{MMES}\rangle  ,
    \label{multipleconvergent}
\end{align}
where the product is evaluated in the reverse order such that $ r = 1 $ is applied first. The full sequence for the adaptive sequential QND measurements is written
\begin{align}
    \vec{\vec{\Delta}} = (\vec{\Delta}^z_1,\vec{\Delta}^x_1,\vec{\Delta}^z_2,\vec{\Delta}^x_2\dots\vec{\Delta}^z_M,\vec{\Delta}^x_M).
    \label{multipleprojsequence}
\end{align}
The two repeat-until-success sequences in the $ z $ and $ x $ basis are repeated until convergence is attained, and defined as obtaining the outcome $ \Delta = 0 $ for the first measurement in each repeat-until-success sequence.  

Here we summarize, for the sake of clarity, the entire
protocol for preparing the MMES using adaptive QND scheme (Fig. \ref{fig3}). The protocol follows the sequence: 
\begin{enumerate}
	%\item The system is prepared in the initial state i.e. $S^x$- polarized state \ref{initialstate}.
	\item Perform the repeat-until-success $\Pi^{(z)}_\Delta$ QND measurement sequence in the $z$ basis.  If $\Delta \ne 0$, then apply unitary $U^{(z)}_\Delta$ as a correction and reapply  $\Pi^{(z)}_\Delta$ until the measurement outcome $\Delta = 0$ is obtained (\ref{projsequence}). 
	\item Do the same as step 1 in the $x$ basis in order to converge towards $\Delta = 0$ measurement outcome.
	%Another QND measurement is performed in the $x$- basis to improve the correlations and entanglement. The projection measurements for the maximal entanglement are corrected.
	\item Repeat steps 1 and 2 until the outcome $\Delta = 0$ is obtained for both on the first measurement for a satisfactory number of cycles (\ref{multipleconvergent}).  
	%We apply the stroboscopic sequence as in step (ii), (iii) and (iv) to improve the overall probability and fidelity of obtaining the macroscopic maximally entangled state. 
\end{enumerate}
The above sequence, using adaptive QND, deterministically converges an initial state to a MMES (\ref{transformedsingletstatebec}).

%Physically the process corresponds to perform measurements on the spins that entangle the spin states. If this is not the desired state, one needs to perform rotation in the spin space. Since we are targeting the maximally entangled state, as soon as the desired correlations establish, the measurement process stops else we repeat the whole process. This repetitive measurement process continues till the system converges towards the macroscopic singlet state.

\subsection{The adaptive unitary}
In this section we discuss the choice of unitary rotation that is employed in the repeat-until-success sequence. There is in fact no unique choice for the adaptive unitary and we take advantage of this to choose a convenient form that has a simple experimental implementation.   In order to understand the different choice of the unitary rotation, we first analyze the state,
\begin{align}
| \tilde{\psi}_{\Delta }^c \rangle & = U^{(z)}_{\Delta }\Pi^{(z)}_{\Delta }|\psi_0\rangle.
\label{singlepsicorrect}
\end{align}

The main criterion for the unitary correction is that it maximizes the probability that $\Delta = 0$ is obtained in the next outcome.  As may be seen by the measurement operator (\ref{projzbasis}), there are two outcomes which occur randomly depending on detection count of the photonic outcomes $n_d$.  We assume that $n_d$ is not measurable, since it requires single photon resolution of a bright laser, which is experimentally challenging.  In order to overcome this, we choose a unitary correction that rotates  the state such that it has a significant overlap with the $\Delta = 0$ sector, regardless of the random outcome of $ n_d $ in QND measurements (\ref{projzbasis}). In the previous works of Ref. \cite{behbood2014,behbood2013feedback}, post-selection based on the measurement outcomes was utilized to target singlet states with $\textbf{S} = 0$. Additionally, a feedback mechanism \cite{toth2010generation} was employed to enhance the spin correlations with $\textbf{S} = 0$. Our scheme shares similarities with these coherent feedback techniques, as it aims to create the spin correlations present in a MMES in subsequent rounds of QND measurements in a deterministic sense.

We choose a unitary transformation that is based on a spin rotation
\begin{align}
U^{(z)}_{\Delta } &= e^{iS^y_1\frac{\theta_{\Delta}}{2}}\otimes I_2 . 
\label{correction}
\end{align}
\begin{figure}[t]%
	\includegraphics[width=\linewidth]{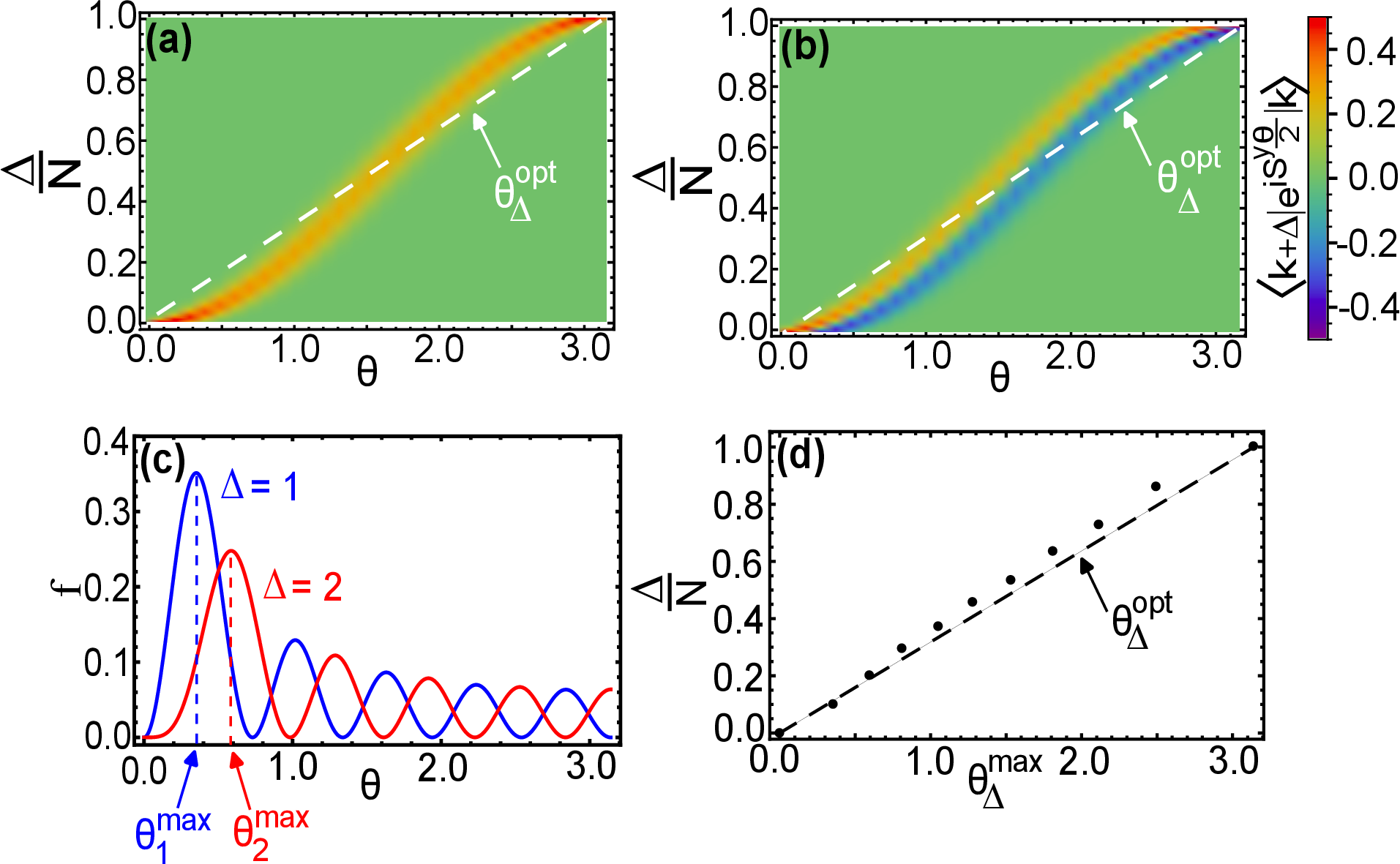}
	\caption{Choice of optimum angle of unitary transformation: Plot of the matrix element $\langle k+\Delta| e^{iS^y\frac{\theta}{2}} |k\rangle$ given in (\ref{matrixelem}) as a function of the angle of unitary rotation $\theta$ in (\ref{correction}) with the measurement outcome $\Delta$ in (\ref{multizseq}) for (a) $k=0$, (b) $k=1$. Total number of the atoms is $N = 150$, (c) Variation of the fidelity (\ref{fid}) with the angle of unitary rotation (\ref{correction}) in the adaptive QND measurement outcomes, $N = 10$.  $\theta^{\text{max}}_{\Delta}$ represents angle of unitary rotation that maximizes the fidelity (\ref{fid}) for a particular measurement outcome $\Delta$,  (d) Plot of maximized angles of unitary rotation for different measurement outcome (\ref{singlepsicorrect}), $N = 10$. The dashed line in (a),(b),(d) depicts the optimized choice of unitary rotation that maximizes the fidelity, which is fitted with the line $\theta^{\text{opt}}_{\Delta}=\pi\frac{\Delta}{N}$.}
	\label{fig33}%
\end{figure}
%
%\begin{align}
%U^{(z)}_{\Delta } &= e^{iS^z_1\frac{\pi}{4}}e^{iS^y_1\frac{\theta}{2}}e^{-iS^z_1\frac{\pi}{4}}\otimes e^{-iS^z_2\frac{\pi}{4}}e^{-iS^y_2\frac{\theta}{2}}e^{iS^z_2\frac{\pi}{4}}\nonumber\\
%& = e^{i(S^x_1+S^x_2)\frac{\theta}{2}}
%\label{correction}
%\end{align}
%
We require a relationship between the measurement outcome $ \Delta $ and the corresponding angle of rotation $\theta$. An adaptive unitary (\ref{correction}) changes the QND measured initial state (\ref{projinitial}) as,
\begin{align}
U^{(z)}_\Delta\Pi_{\Delta}^{(z)} | \psi_0 \rangle & = \sum_{k=0}^{N-\Delta} \sum_{k'=0}^{N} \psi_{k}^{+} \langle k'|e^{iS^y_1\frac{\theta_{\Delta}}{2}}|k + \Delta \rangle |k' \rangle_1 | k  \rangle_2 \nonumber  \\  & + 
\sum_{k=\Delta}^{N} \sum_{k'=0}^{N} \psi_{k}^{-} \langle k '|e^{iS^y_1\frac{\theta_{\Delta}}{2}}|k - \Delta \rangle |k' \rangle_1 | k  \rangle_2 . 
\label{projunitinitial}
\end{align}
%
%\begin{align}
 %   \psi_{k}^{+} & = \frac{1}{2^{ \delta_{\Delta} }}\langle k + \Delta| \langle k  |\psi_0\rangle \nonumber \\
  %  \psi_{k}^{-} & = \frac{(-1)^{(1- \delta_{\Delta})n_d}}{2^{ \delta_{\Delta} }}\langle k - \Delta| \langle k |\psi_0\rangle.
 %   \label{coefftransformedstate}
%\end{align}
We see that the modified state (\ref{projunitinitial}) involves the matrix elements of unitary rotations $  e^{i S^y \theta/2 } $.  

In order to maximize the probability that the outcome $ \Delta = 0 $ in the next measurement is obtained, we require performing a rotation %that the Fock state numbers of the two BECs are equal.  
within the state space that transforms the random projected state to MMES while maintaining the overall coherence and entanglement properties.
Mathematically, it translates to
%Concretely, we require 
maximizing the amplitudes of the terms with $ k' = k $ in (\ref{projunitinitial}), such that the matrix elements $ \langle k | e^{i S^y \theta/2 } | k \pm \Delta \rangle $ have a large value (see Appendix \ref{app:prob} for an explicit expression of the matrix elements).
%in order to get the optimize choice for angle of rotation in the unitary (\ref{correction}).  
%that is used to correct the initial state $|\psi_0\rangle$ to the maximally entangled state (\ref{transformedsingletstatebec}). Clearly, the unitary rotation (\ref{correction}) does not transform the state when the measurement outcome at $\Delta=0$ is encountered. Therefore, one possible choice for angle of rotation $\theta$ dependent on the measurement outcome $\Delta$ is given by
%

Fig. \ref{fig33}(a),(b) shows the plot of the amplitude of matrix element $ \langle k + \Delta | e^{i S^y \theta/2 } | k \rangle $ for two values of $k=0,1$ respectively. We can see that the largest amplitudes occur for a unitary rotation corresponding to a particular outcome $\Delta$ near to the curve
\begin{align}
    \theta_{\Delta} \propto \frac{\Delta}{N}.
    \label{optimumangle}
\end{align}
We see that as $ k $ increases in Fig. \ref{fig33}(a),(b), the region where the matrix elements have a significant magnitude broadens.  

To find the proportionality constant in (\ref{optimumangle}), we analyze the overlap of the transformed state with MMES. %(\ref{matrixelem}) such as $\langle k+\Delta|  e^{iS^y\frac{\theta}{2}} |k\rangle$ and $\langle k-\Delta|  e^{iS^y\frac{\theta}{2}} |k\rangle$ that appear in transformation of unitary operators,
%\begin{align}
 %   U^{(z)}_\Delta \Pi^z_\Delta  U^{(z)\dagger}_\Delta = \tilde{\Pi}^z_\Delta
%\end{align}
%The angle of rotation $\theta$ is chosen in such a way such that the transformed measurement operator $\tilde{\Pi}^z_\Delta$ closely resembles with that of $\Pi^z_0$ and hence, the measured state has the maximum overlap with the maximally entangled state (\ref{eprstate}).
The fidelity of the normalized state (\ref{singlepsicorrect}) with the MMES (\ref{eprstate}) is calculated after the first QND measurement as 
\begin{align}
    f = \frac{|\langle \text{MMES}|\tilde{\psi}_{\Delta}^c \rangle |^2}{\langle \tilde{\psi}_{\Delta}^c|\tilde{\psi}_{\Delta}^c\rangle}.
\label{fid}
\end{align}
Fig. \ref{fig33}(c) shows the variation of the fidelity of the state when the angle of unitary rotation is varied. We can see that the fidelity is maximum for a particular angle of unitary rotation $\theta^{\text{max}}_{\Delta}$. It is clear that the choice of the angle is unique that maximizes the fidelity.

Fig. \ref{fig33}(d) shows the possible choice for the angle of unitary rotation, we see that the largest amplitude occurs near to the line
\begin{align}
\theta^{\text{opt}}_{\Delta}=\pi\frac{\Delta}{N} . 
    \label{optimizedangle}
\end{align}
This corrects the state (\ref{singlepsicorrect}) in such a way that it has a large overlap with the MMES (\ref{eprstate}) in the next round of measurement. We note that it is possible to further improve upon the choice (\ref{optimizedangle}), but we find that this is a simple but effective choice that works for all $ N $.

\section{Performance of the adaptive QND scheme}
\label{sec5}
In order to demonstrate that the MMES is prepared using our protocol, we have performed a numerical analysis to check the effectiveness of the protocol.
%This is done by calculating the density probability distributions in Fock space, probability of obtaining the singlet state in stroboscopic sequence and the overall fidelity of the final projected state.

\subsection{Convergence to desired measurements}
We first examine the probability distribution of the state after one QND measurement and correction step (\ref{multizconvergence}) in the $z$ basis according to the protocol. The probability of a particular sequence is defined by,
\begin{align}
p_{\vec{\Delta}} & =\langle \tilde{\psi}_{\vec{\Delta }} | \tilde{\psi}_{\vec{\Delta }} \rangle \nonumber \\
    & = \langle \psi_0| T^{(z)\dagger}_{\vec{\Delta}} T^{(z)}_{\vec{\Delta}} |\psi_0 \rangle
\label{normprobdelta}
\end{align}
where the normalized state (\ref{multizconvergence}) of the protocol is given by,
\begin{align}
|\psi_{\vec{\Delta }} \rangle & = \frac{|\tilde{\psi}_{\vec{\Delta }}  \rangle}{{\sqrt{\langle \tilde{\psi}_{\vec{\Delta }} |\tilde{\psi}_{\vec{\Delta }} \rangle}}}.
\label{normalizedstateadaptiveqnd}
\end{align}
\begin{figure}[t]%
	\includegraphics[width=\linewidth]{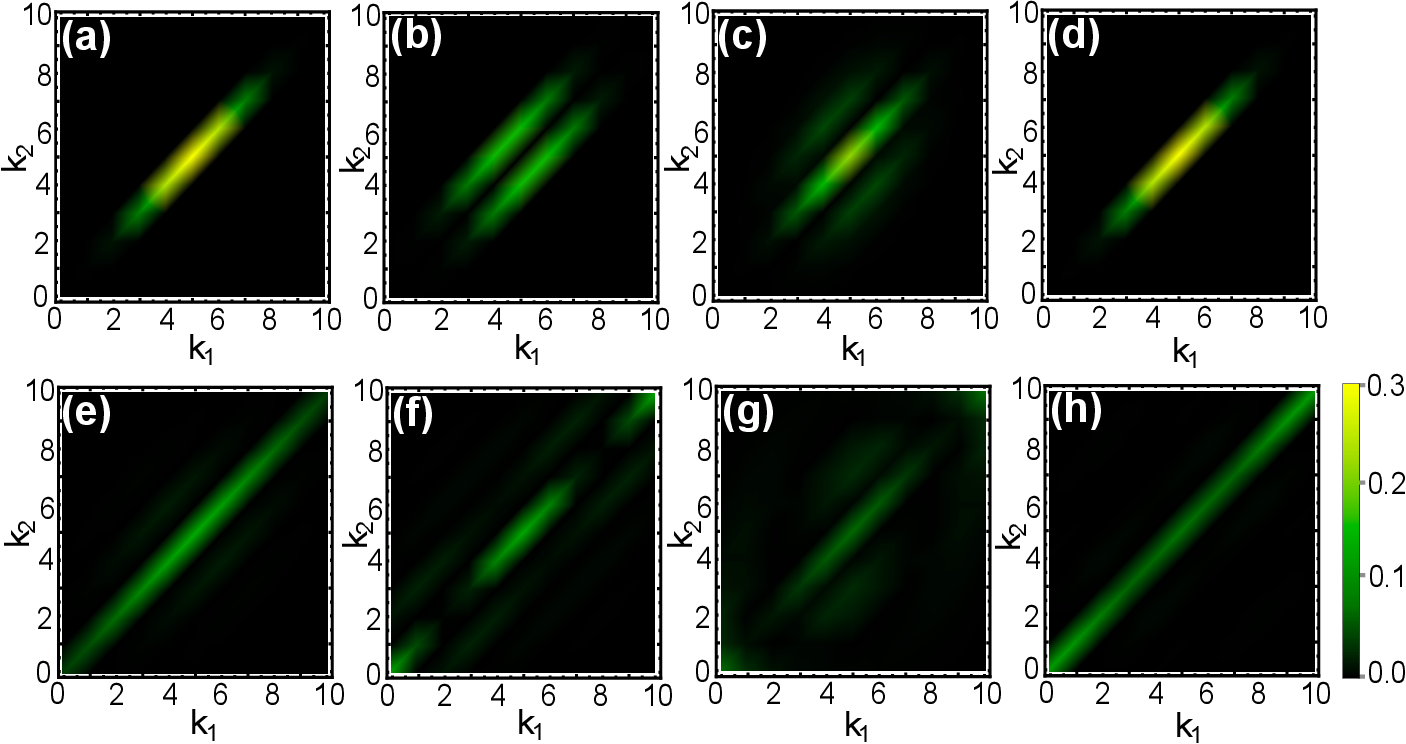}
	\caption{Plots for the probability distribution (\ref{probdistribution}) of the initial state (\ref{initialstatexx}) after sequential adaptive QND measurement (\ref{multipleconvergent}) in the $z$ and subsequently in the $x$ basis for operators (a)$ \Pi^z_0 $, (b)$\Pi^z_1$, (c)$U^z_1\Pi^z_1$, (d)$\Pi^z_0U^z_1\Pi^z_1$, (e)$\Pi^x_0\Pi^z_0$, (f)$\Pi^x_2\Pi^z_0$, (g)$U^x_2\Pi^x_2\Pi^z_0$ and (h)$\Pi^x_{0}U^x_2\Pi^x_2\Pi^z_0$. The number of atoms in each ensemble is $N = 10$.
	}
	\label{fig4}%
\end{figure}

We consider the initial state of the two atomic ensembles to be $S^x$-polarized state,
\begin{align}
    |\psi_0\rangle &=\Big|\frac{\pi}{2},0\Big\rangle \Big\rangle_1 \Big|\frac{\pi}{2},0\Big\rangle \Big\rangle_2  \nonumber \\
    & =\frac{1}{2^N} \sum_{k_1,k_2 = 0}^{N} \sqrt{{N \choose k_1} {N \choose k_2}} |k_1 , k_2\rangle.
		\label{initialstatexx}
\end{align}
%where the binomial factors in (\ref{initialstatexx}) can be approximated by a Gaussian for $N \gg 1$ as,
%$\begin{align}
  %  \frac{1}{2^N} \sqrt{{N \choose k} }\approx \sqrt{\frac{2}{N\pi}} e^{-\frac{2}{N}(k-\frac{N}{2})^2}
%\end{align}
%The effect of QND measurement on the initial state (\ref{initialstatexx}) corresponding to outcome $\Delta =0$  is described as,
%
The operator $ T_{\vec{\Delta}}^{(z)} $ applied on the initial state (\ref{initialstatexx}) produces correlations between the BECs in the $z$ basis.  In the case of obtaining $ \Delta = 0 $ outcome on the first measurement, the state that is obtained is
\begin{align}
   \Pi^z_0 |\psi_0\rangle & = \frac{1}{2^N} \sum_{k = 0}^{N} { N \choose k }  |k , k\rangle \nonumber\\
   & = \sum_{k = 0}^{N} \sqrt{p_{0}(k,k)} |k , k\rangle ,
   \label{deltazero}
\end{align}
where $p_{\Delta}(k_1,k_2) = |\langle k_1,k_2|\Pi^z_{\Delta} |\psi_0\rangle|^2$ is the probability of the measured state for a particular outcome $\Delta$ in the Fock basis. The outcome $\Delta=0$ signifies the MMES-like correlations (\ref{eprstate}). 

In general, for a random measurement sequence (\ref{multipleconvergent}),
the probability distribution in the Fock states is described as,
\begin{align}
    p_{\vec{\vec{\Delta}}}(k_1,k_2) & = | \langle k_1,k_2|T^{(x)}_{\vec{\Delta}^x}  T^{(z)}_{\vec{\Delta}^z}  |\psi_0 \rangle|^2.
    \label{probdistribution}
\end{align}
In Fig. \ref{fig4} we plot the probability distribution of the state (\ref{probdistribution}) after performing QND measurement and correction operations in the $z$ and $x$ basis respectively. In Fig. \ref{fig4}(a)-(d) we show the probability distributions for one measurement and unitary correction sequence in the $z$ basis. In Fig. \ref{fig4}(a) we see that the probability distribution for $\Delta^z = 0$ is correlated along $k_1 = k_2$ in the Fock state space of two ensembles and it resembles as that of the MMES distribution (\ref{eprstate}). It is however not the MMES because of the binomial factors in (\ref{deltazero}). For the projection outcomes $\Delta^z = 1$ in Fig. \ref{fig4}(b), we see the offset in Fock state probability distribution with $k_2=k_1\pm \Delta^z$ according to the definition of operator (\ref{projzbasis}). By applying a unitary correction (\ref{correction}), mostly the probability distribution is restored along the diagonal as shown in Fig. \ref{fig4}(c), such that in the subsequent measurement there is a high probability of obtaining $\Delta^z = 0$ in Fig. \ref{fig4}(d).

Fig. \ref{fig4}(e)-(h) show the effect of another application of the sequence of QND measurements (\ref{projsequence}), where the basis is changed from $z$ to $x$. Correlations are further improved in Fig. \ref{fig4}(e) because of suppression of the binomial factors (\ref{deltazero}). Unlike Fig. \ref{fig4}(b), we observe weaker offsets in the Fock state space as it is clear from Fig. \ref{fig4}(g)-(h). It is because of the fact that in subsequent QND measurement and corrections, stronger spin correlations are developed only for the MMES. Hence, the probability of obtaining the prepared state in other measurement outcomes, such as $\Delta^x \neq 0$, is less likely and the probability distribution converges solely towards that of the MMES in Fig. \ref{fig4}(f)
which implies the deterministic preparation of an initial state from the scheme.

\begin{figure}[t]%
	\includegraphics[width=\linewidth]{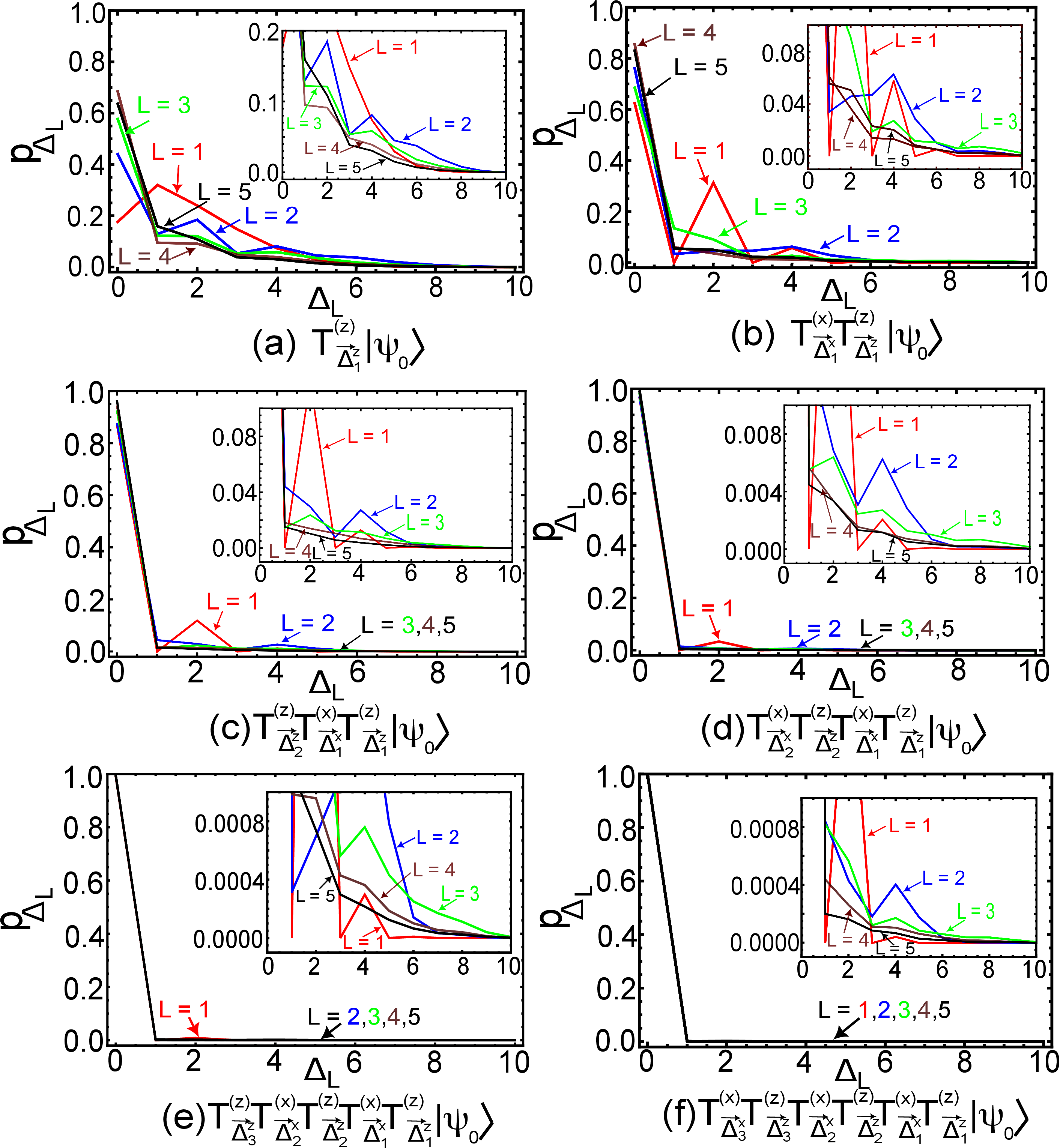}
	\caption{Marginal probability (\ref{marginalprobability}) for different measurement outcomes in sequential adaptive QND measurement (\ref{multipleconvergent}) is shown for two atomic ensembles prepared in $S^x$-polarized state (\ref{initialstatexx}). Convergence is attained for measurement outcome $\Delta = 0$ after three rounds of iterations. A zoomed in plot is shown in the inset for better visibility of the probability values and its convergence in a sequence.
The number of atoms in each ensemble is $N = 10$.	}
	\label{fig5}%
\end{figure}

\subsection{Probability distribution}
We now turn to the probability (\ref{normprobdelta}) of the various measurement outcomes in the protocol, shown in Fig. \ref{fig5}.
We define the marginal probability distribution of obtaining the measurement outcomes in a particular sequence (\ref{multizseq}) as,
\begin{align}
    p_{\Delta_L} = \sum_{\Delta_1,\Delta_2\dots\Delta_{L-1}}p_{\vec{\Delta}}. 
    \label{marginalprobability}
\end{align}
%where the probability $p_{\vec{\Delta}} $ in general is,
%\begin{align}
   % p_{\vec{\Delta}} \equiv p(\Delta_1,\Delta_2\dots\Delta_{L})
%\end{align}
The marginal probability gives the total probability of obtaining an outcome $\Delta_L$ in a sequence of $L$ measurements (\ref{multizseq}). This gives the probability of obtaining an outcome $\Delta_L$ in a sequence, regardless of the previous measurement outcomes.  

In Fig. \ref{fig5} we have plotted the marginal probabilities for various levels of iteration for different measurement sequences (\ref{projsequence}).  Fig. \ref{fig5}(a) shows a single $z$ basis measurement sequence. As we can see, the marginal probability for the initial state is generally largest for the outcome $\Delta =0 $ and the probability decreases for other outcomes $\Delta \neq 0 $. The probability to obtain the MMES increases with larger numbers of measurements ($L=5$) in a sequence. Fig. \ref{fig5}(b) shows an $x$ basis sequence after an initial measurement sequence in the $z$ basis, where the final outcome was $\Delta^z = 0$. The probability for obtaining the MMES increases successively with the measurement sequences as compared to Fig. \ref{fig5}(a) and hence, other probabilities corresponding to the measurement outcomes $\Delta \neq 0$ are suppressed further. Similarly, Fig. \ref{fig5}(c)-(d) show another $z$ and $x$ basis measurement sequences respectively ($M = 2$) after the first $z$ and $x$ basis sequence ($M = 1$), in this case the state converges to the MMES at a faster rate. The state is prepared in the measurement outcome $\Delta = 0$ with almost unit probability and the other measurement outcomes $\Delta \neq 0$ occur with low probability. Finally, Fig. \ref{fig5}(e)-(f) best describes the overall performance of the protocol as the probability of obtaining the outcome $\Delta = 0$ is dominant in the subsequent QND measurements in the $z$ and $x$ basis respectively ($M=3$), and the MMES is prepared with nearly $100\%$ success with very little contribution from the other measurements because of the increasing spin correlations. This shows that the MMES can be prepared in a deterministic way.

\subsection{Success probability}
In the previous section, we have seen that in the sequential adaptive QND measurements, the probability of obtaining $\Delta \neq 0$ measurement outcomes is low and the system is prepared deterministically in the MMES with outcome $\Delta=0$. We define the success probability for obtaining the MMES as a sum of the probability of all the measured states in a QND measurement sequence with measurements that end with $\Delta=0$:
\begin{align}
p_{\text{suc}} & = p_{\Delta_L = 0 }\nonumber\\
& =\sum_{\Delta_L \epsilon \{0\}} \sum_{\Delta_1,\Delta_2\dots\Delta_{L-1}}p_{\vec{\Delta}}
\label{successprob}
\end{align}
Fig. \ref{fig6} shows the success probability of obtaining the MMES in our protocol for various levels of iteration.
We see that after a single $z$ basis measurement sequence, (e.g. the $M=1 $ case), the success probability increases monotonically, as expected, although it is not sufficient to drive towards a perfect MMES as other measurement outcomes are still possible (see also Fig. \ref{fig5}(a)). Another measurement sequence in $x$ basis leads to enhanced spin correlations and an increased probability for obtaining $\Delta = 0$ outcome and hence, it shows better success probability. Similarly in the next round of measurements in the $z$ and $x$ basis, i.e. $M=2$, near unit success probability is achieved. After three rounds of measurements ($M=3$), the success probability of obtaining the MMES is close to unity. The convergence to unit probability is shown in the inset for better clarity. 
%\begin{figure}[t]%
%	\includegraphics[width=\linewidth]{Fig6new.eps}
%	\caption{Success probability (\ref{successprob}) for obtaining the macroscopic MMES after stroboscopic QND measurement and correction (\ref{normprobdelta}) for (a) $m =1$, (b) $m =2$ and (c) $m =3$. It shows the convergence to desired state after each pulse in $z$-basis and $x$-basis. The number of atoms in each ensemble is $N = 10$.
%	}
%	\label{fig6}
%
%\end{figure}
\begin{figure}[t]%
	\includegraphics[width=\linewidth]{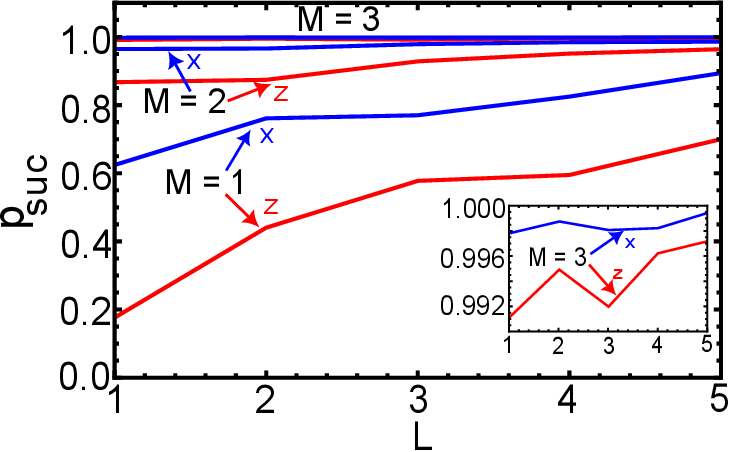}
	\caption{Success probability (\ref{successprob}) for obtaining the MMES after sequential adaptive QND measurement (\ref{multipleconvergent}) for $M =1,2,3$ is plotted. It shows the convergence to the desired state after each measurement in the $z$ and $x$ basis. A zoomed in plot is shown for $M=3$ in the inset. The number of atoms in each ensemble is $N = 10$.
	}
	\label{fig6}
\end{figure}

\subsection{Fidelity calculation}
Finally, we calculate the fidelity of the final state obtained from the protocol.
The fidelity of the normalized state (\ref{multipleconvergent}) with respect to the MMES in an adaptive QND measurement sequence (\ref{multipleprojsequence}) is calculated as
\begin{align}
	F_{\vec{\vec{\Delta}}} & = \frac{| \langle \text{MMES} | \tilde{\psi}_{\vec{\vec{\Delta}} }^f  \rangle |^2}{\langle \tilde{\psi}_{\vec{\vec{\Delta}} }^f | \tilde{\psi}_{\vec{\vec{\Delta}} }^f \rangle}.
 %\nonumber \\
% & = | \langle \text{MMES} | \prod_{r=1}^{M}(T_{\vec{\Delta}_r^x}^{(x)}T_{\vec{\Delta}_r^z}^{(z)})  |\psi_0 \rangle |^2,
\end{align}
We also define the fidelity over all possible outcomes, the average fidelity is calculated as,
\begin{align}
    F_{\text{avg}} & = \sum_{\vec{\vec{\Delta}}} p_{\vec{\vec{\Delta}}}  F_{\vec{\vec{\Delta}}}
    \nonumber \\
& = | \langle \text{MMES} | \prod_{r=1}^{M}(T_{\vec{\Delta}_r^x}^{(x)}T_{\vec{\Delta}_r^z}^{(z)})  |\psi_0 \rangle |^2,
    \label{avgfidelity}
\end{align}
where the probability of a state in a particular sequence is 
\begin{align}
    p_{\vec{\vec{\Delta}}} = \langle \tilde{\psi}_{\vec{\vec{\Delta}} }^f | \tilde{\psi}_{\vec{\vec{\Delta}} }^f \rangle.
\end{align}
Fig. \ref{fig7} shows the average fidelity for obtaining the MMES for our protocol (\ref{multipleconvergent}). In the first $z$ basis measurement, the average fidelity is low, and it increases with the number of measurements made in a sequence. An $x$ basis measurement after an initial measurement sequence in the $z$ basis in the final outcome $\Delta^z = 0$ improves the average fidelity as the probability for obtaining the MMES increases. In the next round of measurements in the $z$ and $x$ basis, i.e. $M=2,3$, the average fidelity increases to unity implying perfect preparation of the MMES only.
 %, as obvious. Thus, the initial state of the system converges towards the MMES at measurement outcome $\Delta = 0$ with stronger spin-correlations, unit probability and unit fidelity. 
 %
%\begin{figure}[t]%
%	\includegraphics[width=\linewidth]{Fig7new.eps}
%	\caption{Fidelity of the state for different projection outcomes (\ref{normprobdelta}) in stroboscopic QND measurement and correction (\ref{multipleconvergent}). Convergence is attained after three rounds of measurements for $\Delta = 0$ projection outcome.
%The number of atoms in each ensemble is $N = 10$.}
%	\label{fig7}%
%\end{figure}
%
 %
\begin{figure}[t]%
	\includegraphics[width=\linewidth]{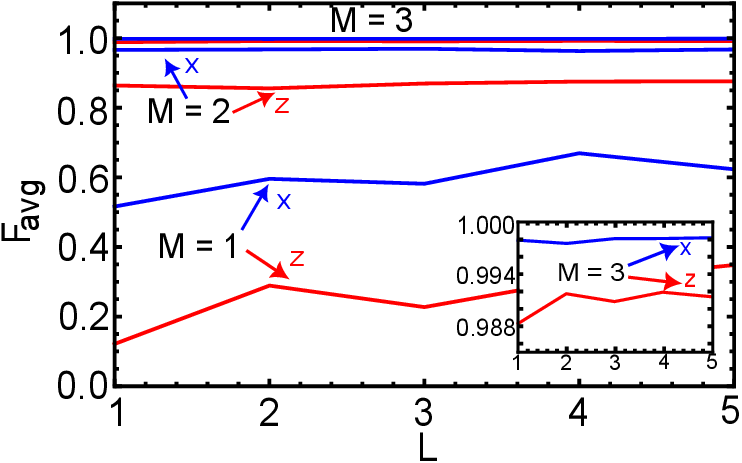}
	\caption{Average fidelity (\ref{avgfidelity}) of the initial state (\ref{initialstatexx}) for different measurement outcomes is calculated in adaptive QND measurement (\ref{multipleconvergent}) for $M = 1,2,3$. Convergence is attained after three rounds of measurements. A zoomed in plot is shown for $M=3$ in the inset. The number of atoms in each ensemble is $N = 10$.
 }
	\label{fig7}%
\end{figure}

\section{Effect of imperfections}
%\section{Effect of imperfections}
\label{sec6}
Here we discuss the performance of the protocol with the possible sources of decoherence included. Specifically, we discuss the effect of the atom number fluctuations and the initial ensembles prepared in a maximally mixed state.
\subsection{Initial maximally mixed state}
\label{sec6A}
We consider the initial state of the system to be a maximally mixed state described by
\begin{align}
    %\rho_0 = \frac{\mathds{1}\otimes\mathds{1}}{(N+1)^2}
    \rho_0 = \frac{I_1 \otimes I_2}{(N+1)^2},
    \label{thermalstate}
\end{align}
where $I_j$ is the identity matrix in the Hilbert space for the $j^{\text{th}}$ ensemble. The procedure is identical to the pure state calculation performed earlier using the equations (\ref{successprob})-(\ref{avgfidelity}).
The action of the first QND measurement in $z$ spin basis (\ref{projsequence}) transforms the density matrix,
%\begin{align}
  %  \rho_0 \to T_{\vec{\Delta}}^{(z)}  \rho_0 T_{\vec{\Delta}}^{(z) \dagger}
%\end{align}
such that the protocol leads to the convergence to the MMES state
\begin{align}
    \rho_{\vec{\vec{\Delta}}} = \prod_{r=1}^{M}(T_{\vec{\Delta}_r^x}^{(x)}T_{\vec{\Delta}_r^z}^{(z)}) \rho_0 (T_{\vec{\Delta}_r^x}^{(x)}T_{\vec{\Delta}_r^z}^{(z)})^{\dagger}.
\end{align}
The fidelity is calculated as
\begin{align}
F_{\vec{\vec{\Delta}}}^{\text{mixed}} = \frac{ \langle \text{MMES} |\rho_{\vec{\vec{\Delta}}}| \text{MMES}\rangle}{\text{Tr}(\rho_{\vec{\vec{\Delta}}})},
\end{align}
and the average fidelity is expressed as
\begin{align}
    F_{\text{avg}}^{\text{mixed}} = \sum_{\vec{\vec{\Delta}}} p_{\vec{\vec{\Delta}}}  F_{\vec{\vec{\Delta}}}^{\text{mixed}}
\end{align}
where the probability of a particular sequence is 
\begin{align}
   p_{\vec{\vec{\Delta}}}  = \text{Tr}(\rho_{\vec{\vec{\Delta}}}).
\end{align}
%
%\begin{align}
  %   \prod_{r=1}^{M}(T_{\vec{\Delta}_r^x}^{(x)}T_{\vec{\Delta}_r^z}^{(z)}) \rho_0 (T_{\vec{\Delta}_r^x}^{(x)}T_{\vec{\Delta}_r^z}^{(z)})^{\dagger}\xrightarrow{M \rightarrow \infty }
  %  |\text{MMES}\rangle \langle \text{MMES}|
%\end{align}
%
Figure \ref{Fig8}(a)-(b) show the success probability and average fidelity using the above procedure. In Fig. \ref{Fig8}(a), the probability of achieving the MMES increases in a similar sequential manner to that observed in Fig. \ref{fig6}. However, in Fig. \ref{Fig8}(b), the fidelity of the prepared state is initially lower compared to that of Fig. \ref{fig7}. This discrepancy is in contrast to the initial spin coherent states (\ref{initialstatexx}) that collapses to the state (\ref{deltazero}) in a single QND measurement. This does not happen with the initial state in (\ref{thermalstate}) and the convergence rate to the MMES is slower in the initial rounds of adaptive QND measurements. After two rounds, $M=2$, this converges to the MMES rapidly as seen in Fig. \ref{Fig8}(b). Since the initial maximally mixed state lacks coherence, and any coherence is entirely produced by the subsequent QND and unitary rotations, its evolution within the adaptive QND scheme towards the MMES shows the overall robustness of the protocol.
\begin{figure}[t]%
\includegraphics[width=\linewidth]{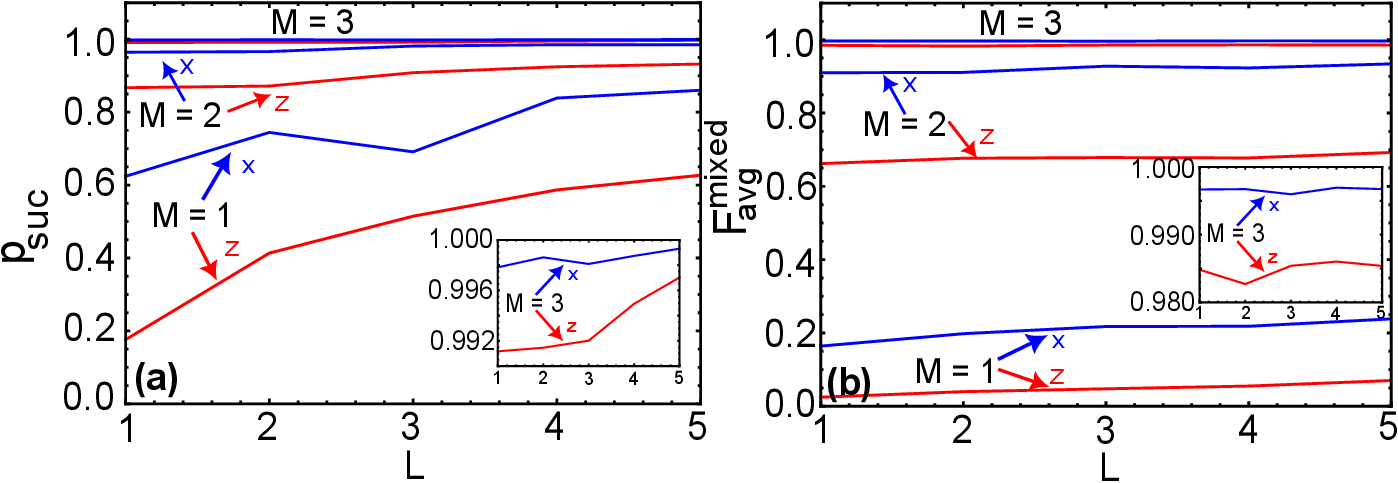}
	\caption{The success probability (\ref{successprob}) and average fidelity (\ref{avgfidelity}) of the initial mixed state (\ref{thermalstate}) for obtaining the MMES after sequential adaptive QND measurement (\ref{multipleconvergent}) for $M =1,2,3$ is plotted. A zoomed in plot is shown for $M=3$ in the inset. 
 The number of atoms in each ensemble is $N = 10$.}
	\label{Fig8}
\end{figure}
\subsection{Atom number fluctuations}
\label{sec6B}
In a typical experimental setup, BECs are not prepared with fixed atom number. In this section, we discuss the effect of the atom number fluctuations when both BECs are prepared as the statistical distribution of different atom numbers $N$. Here we consider an initial state that is a mixed state of various atom numbers according to
\begin{align}
    \rho_0 = \sum_{N_1 N_2} p(N_1) p(N_2) |\psi_0 \rangle_{N_1,N_2} \langle \psi_0 |_{N_1,N_2}.
\end{align}
Here, $\rho_0$ is the initial state density matrix and the probabilities $p(N_l)$ are taken to be Gaussian distributions.
%The action of first QND measurement in $z$ spin basis (\ref{projsequence}) transforms the density matrix,
%\begin{align}
  %  \rho_0 \to T_{\vec{\Delta}}^{(z)}  \rho_0 T_{\vec{\Delta}}^{(z) \dagger}
%\end{align}
%such that the protocol leads to the convergence to the MMES state,
%\begin{align}
 %   \rho_{\vec{\vec{\Delta}}} = \prod_{r=1}^{M}(T_{\vec{\Delta}_r^x}^{(x)}T_{\vec{\Delta}_r^z}^{(z)}) \rho_0 (T_{\vec{\Delta}_r^x}^{(x)}T_{\vec{\Delta}_r^z}^{(z)})^{\dagger} 
%\end{align}
%and the fidelity is calculated as,
%\begin{align}
  %  F_{\text{avgfluc}} = Tr(\rho_{\vec{\vec{\Delta}}})
%\end{align}
%\begin{align}
 %   \prod_{r=1}^{M}(T_{\vec{\Delta}_r^x}^{(x)}T_{\vec{\Delta}_r^z}^{(z)}) \rho_0 (T_{\vec{\Delta}_r^x}^{(x)}T_{\vec{\Delta}_r^z}^{(z)})^{\dagger} \xrightarrow{M \rightarrow \infty }
   % |\text{MMES}\rangle \langle \text{MMES}|
%\end{align}
%
When accounting for the fluctuations in the atom number for $j^{\text{th}}$ atomic ensemble, denoted as $N_j\pm\delta N_j$, we observe different potential effects that could impact the performance of the protocol. Firstly, the QND interaction and the adaptive unitary rotations themselves do not depend upon the atom number $N$. 
%The particular measurement sequence follows $\approx (N_1+1)\otimes(N_2+1)$ dependence with $N_j$. 
All the operations are atom number conserving. Hence the overall protocol is not modified with the change in $N_j$. For this reason, we have analyzed each particle number sector separately and calculate the fidelity for each sector with the initial state (\ref{initialstatexx}). The only thing that is sensitive is the angle of rotation, which involves $N_j$ dependence. %But even this is a weak dependence due to the formula that you give. %You should talk about the $N$-dependence of the T operators and say that parts like the QND measurement and unitary rotation operator does not change with $N$. You should also say that .   
%However, there is a dependence on the atom number for the adaptive unitary rotation (\ref{optimizedangle}). 
%Considering equal atom number and atom fluctuations for simplicity, 
With fluctuations in the atom number $N'=N\pm \delta N$, the unitary rotation angle is modified to 
\begin{align}
    \theta'^{\text{opt}}_{\Delta} \approx \theta^{\text{opt}}_{\Delta} \Big(1\pm \frac{\delta N}{N} \Big)  .
    \label{optimizedangleatomfluc}
\end{align}
In the recent experimental work of Ref. \cite{becwithatomfluc}, it was found that $\frac{\delta N}{N}$ ranges between $10$-$15\%$.
\begin{figure}[t]%
\includegraphics[width=\linewidth]{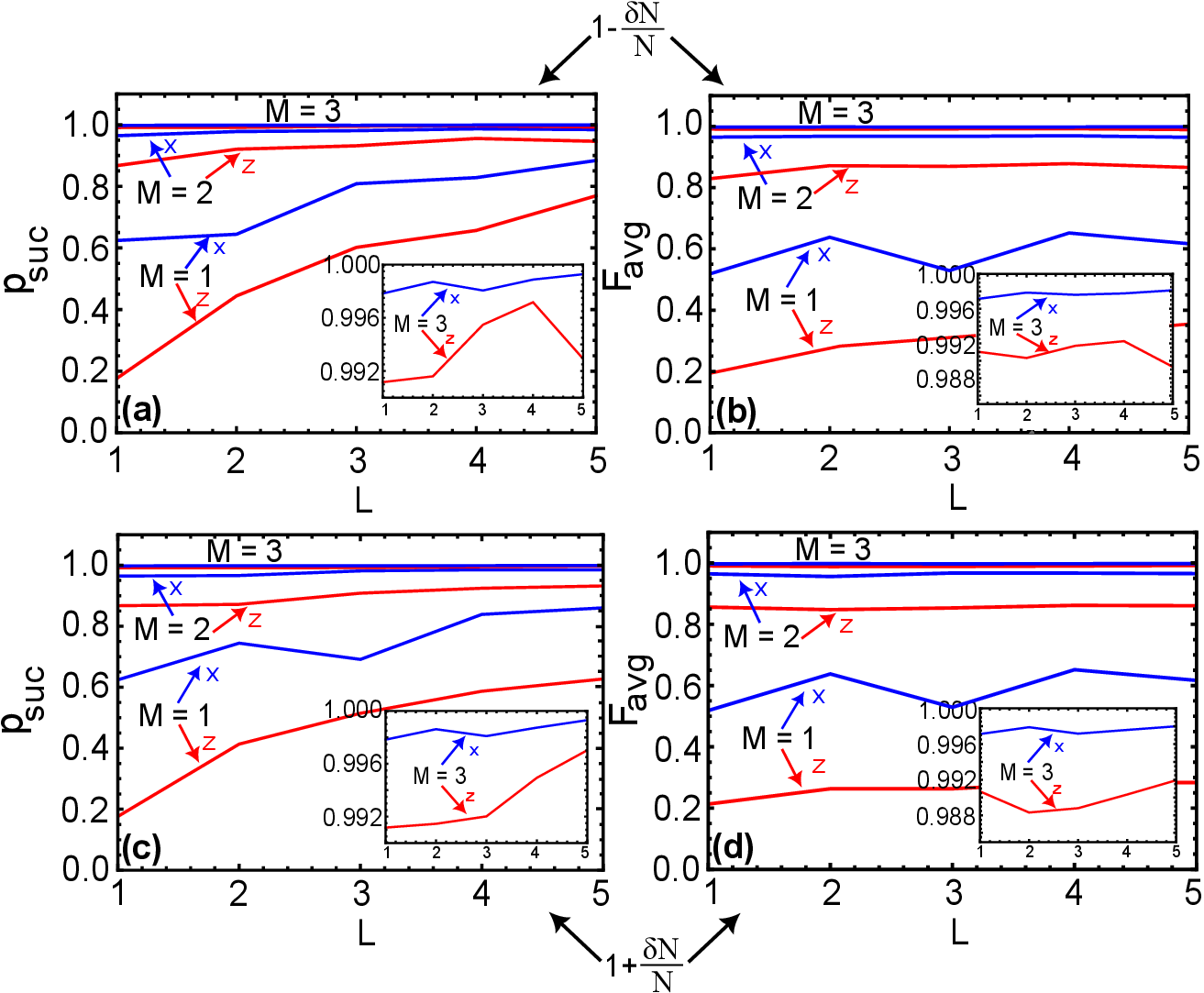}
	\caption{Effect of the atom number fluctuations on the protocol: the success probability (\ref{successprob}) and average fidelity (\ref{avgfidelity}) of the initial state (\ref{initialstatexx}) for obtaining the MMES after sequential adaptive QND measurement (\ref{multipleconvergent}) for $M =1,2,3$ is plotted. The corrective unitary rotation is applied considering the atom number variation in the range (\ref{optimizedangleatomfluc}). %It shows the convergence to the desired state after each measurement in the $z$ and $x$ basis. A zoomed in plot is shown for $M=3$ in the inset. 
 The number of atoms in each ensemble is $N = 10$ and $\frac{\delta N}{N} = 14\%$.}
	\label{Fig9}
\end{figure}
\newline
In Fig. \ref{Fig9} we have analyzed the effect of atom fluctuations on the protocol. Fig. \ref{Fig9}(a)-(d) shows the success probability and average fidelity for obtaining the MMES for various levels of iterations when $N'=N  \pm \delta N$. Within a sequence, it follows a similar trend as observed with $\delta N =0$ case in Fig. \ref{fig6} and Fig. \ref{fig7}. Relative atom number fluctuations result in a random choice to the applied unitary corrections (\ref{optimizedangleatomfluc}), and hence, it leads to a non-monotonic convergence to the MMES. Nevertheless the procedure converges to a MMES and it is clear that rotation with a sub optimal angle does not significantly affect the convergence of the protocol, implying the relative insensitivity of the atom number fluctuations  on the scheme.
%, as in unitary rotation, within the possible extreme values. 

%

\section{Experimental implementation}
\label{sec7}
There are various physical systems that can be employed to implement the protocol and physical operations (Fig. \ref{fig1}) described in the previous sections. One such system involves the use of hyperfine states of $^{87}$Rb atoms, specifically $F=1, m_F=-1$ and $F=2,m_F=1$, which form a two-level atom system. Such a system of collection of indistinguishable atoms can be implemented using either hot or cold atomic ensembles or BECs \cite{timquantumoptics2020}. To prepare two atomic systems in the entangled state, QND measurements are employed depending on the photonic detection outcomes. %Recent advancements have provided efficient one-photon count detectors. All the photonic measurements must be rescaled to account for any zero photon count (dark noise) to avoid the false measurement signal. 
QND measurements have proven to be an effective method for inducing entanglement generation between atomic ensembles \cite{kuzmich2000generation,julsgaard2001experimental}. The QND scheme that we employ features a remarkably versatile geometry, enabling the measurement of physical qubits using addressable optical modes. With the Mach-Zehnder configuration, we can achieve entanglement between highly separated qubits, even when the line of sight is obstructed. This unique capability enhances the potential for robust and scalable entanglement generation. In Sec. \ref{sec2}, we have summarized the formalism required to realize the QND measurement operator. For a particular QND measurement, the outcome is a random state and a sequence of QND pulses leads to a stochastic evolution of the system. By applying the conditional unitary operations, the desired state corresponding to the particular measurement outcome can be generated. The conditional unitary rotations are controlled by resonantly driving the transitions between the two states using a laser of appropriate resonant frequency. 
%In the conclusion section, we discuss the possible sources of practical errors in implementing QND measurements and performing photonic detection within limited number resolution that explains the overall effectiveness of the protocol.

\section{Summary and Conclusions}
\label{sec8}
In this paper we have introduced an adaptive QND scheme to generate the MMES between two atomic ensembles. The state is equivalent to a singlet state formed from two macroscopic spins, with total angular momentum zero, up to a local basis transform. Using the basic properties of the singlet state, we have proposed a protocol that can be implemented using QND measurements with adaptive unitary corrections and converges towards the MMES in a deterministic way. Our scheme is experimentally viable in the sense that it does not use complex operations such as transformations on individual atoms, and only involves collective spin operations, projective measurements and local unitary rotations. In order to check the efficiency of the scheme so as to converge the system towards MMES, we have calculated the fidelity, and the success probability to achieve target state after multiple rounds of measurements and corrections in a sequence. We observe that the probability and fidelity of obtaining the desired state increases in subsequent measurements. We have also checked the probability distribution of the measured state in Fock space and confirmed that it matches with spin correlations of the MMES.

Maximally entangled states find number of important applications in quantum information tasks as these serve as resource states for various quantum protocols. In Ref. \cite{julsgaard2001experimental}, generation of two-mode squeezed states (TMSS) was demonstrated under Holstein Primakoff short interaction time regime in two separate gas cells. Here it is important to understand the difference between TMSS and MMES. The amount of entanglement, as calculated by von Neumann entropy, in a TMSS is $\cosh^2r\text{log}_2 (\cosh^2r)-\sinh^2r\text{log}_2 (\sinh^2r)$, \cite{braunstein2005quantum,kitzinger2020}, where $r$ is the squeezing parameter. Typically the squeezing parameter is in the region of $ r \approx 1 $, hence the amount of entanglement is of the order unity \cite{julsgaard2001experimental,kotler2021direct}. Meanwhile, the value for a MMES between two ensembles of dimension $N$ is $\text{log}_2 (N+1)$ \cite{byrnes2013fractality}. This illustrates that the MMES possesses much more entanglement than in the TMSS. Moreover, in the MMES, the linear combination of all spin observables show correlations (or anti-correlations) \cite{kitzinger2020}, while in TMSS, only few spin observables are correlated (or anti-correlated). Our work provides a simple yet powerful method for producing a MMES, and improves upon previous methods \cite{behbood2014,behbood2013real} which rely upon postselection. In addition, we have not preformed any approximation to spin variables in our calculations and have considered the spins in an exact way. The protocol works regardless of the initial state but we have considered the state that has the largest fidelity with the MMES, namely two spin coherent states polarized in the $x$-direction. 

%In the calculations, we have shown the numerical results for atom number $N = 10$. For indistinguishable atom spins, the dimension of the full Hilbert space for two atomic ensembles grows as $(N+1)^2$, the largest system that we could simulate within a reasonable time was $N = 10$. We expect that our scheme works for a larger number of spins $N$. The major challenge is to evaluate matrix multiplications due to the unitary transformations, which have a dimension $(N+1)^2\times(N+1)^2$. Since the number of possible measurement outcomes grows linearly with $N$,  due to the larger Hilbert space that the state must traverse during the evolution. 

% With the larger system size, one expects that a larger number of adaptive QND measurements would be required for convergence as the number of possible measurement outcomes grows linearly with $N$. For indistinguishable atom spins, the dimension of the full Hilbert space for two atomic ensembles grows as $(N+1)^2$. In practical settings we expect that our scheme can be implemented with high fidelity. 

An important topic is how robust our scheme is in the presence of experimental imperfections.  In Sec. VI we performed two such case studies, of starting in a completely mixed state, and studying the effect of atomic number fluctuations.  We have found that our scheme is very robust and converges to the MMES despite the presence of imperfections. In addition to this,
%In this paper we have focused upon introducing the protocol in an idealized setting and experimental imperfections such as decoherence were not considered.  While we have not performed an explicit calculation including decoherence effects, 
we have made several studies of the effects of QND measurements under decoherence previously \cite{gao2022decoherence,ilookeke2023,gao2023}, which gives positive expectations of the performance of the current scheme. We briefly comment on prospects in this regard.  
We first point out that QND measurements have been shown to be remarkably robust against photon loss.  In a previous work, decoherence effects on QND measurements were studied and shown that as long as the QND interaction times are in the short-time regime (as is the case with the measurements considered in this paper),
decoherence on the atoms can be well-controlled \cite{gao2022decoherence}.

Another technical challenge  is the imperfect photonic resolution at the detectors. The primary effect of imperfect detector efficiency $\eta$ is to reduce the average photon number $\alpha$ by an amount proportional to detection efficiency, i.e. $\alpha \rightarrow \alpha\sqrt{\eta}$, and modify the photon counts $ n_c, n_d $  in Eq. (\ref{cfunc}). 
 %As a result, it masks the number of photons at the outputs of the detector so that the actual value is never known, thereby introducing randomness in the readout.  
Under such a replacement, the general form of Eq. (\ref{cfunc}) remains, however, of the same form, which suggests that the impact on entanglement generation itself may be small.  %It will however produce an effective noise in the estimate of measurement outcome $ \Delta $, which can affect the convergence towards the MMES, such that the perfect stability that is seen in Fig. \ref{fig5} will be disrupted. 
In practice such imperfect photon detection may introduce further noise \cite{ilookeke2023}, but the robustness in our scheme is provided by the fact that only $\Delta = 0$ is the convergence point.  This corresponds to $n_d=0$ according to Eq. (\ref{cfunc}), and the precise value of $n_c$ itself does not matter.  Hence, as long as the phase in the interferometer is unaffected, the imperfect photonic detection at the $n_c$ detector does not affect the protocol. 
%However, the convergence point is centred around $n_d=0$ outcome which itself is not affected by the imperfect photon number resolution in our case. On the other hand, photon loss in the path beam at modes $a_1,a_2$ has the potential to introduce a relative phase in the coherent light. This finite phase might lead to an imperfect cancellation of interference effect such that a non-zero photon is detected in the $d$-mode. However, practical photo detectors with efficiency as good as 85\% does not give appreciable change in the photonic counts.
In the case of experimental realization with a bright laser source, limited photon-number resolution therefore should not significantly affect the fixed point at $n_d=0$ and hence the convergence to the maximally entangled state is still expected to take place. 
%The practical inefficiency of the photon detectors does not affect the convergence point and still the protocol can be used to prepare the MMES deterministically as our protocol is not sensitive to counting photon for each measurement outcome. 
The critical part of the protocol is to effectively distinguish the single measurement outcome at $n_d=0$ from other possible outcomes. As shown in Sec. \ref{sec6B}, the protocol is not extremely sensitive to the particular rotation that is performed for $ \Delta \ne 0 $, and affects the convergence speed moderately but not the asymptotic state.

Another potential source of decoherence is the spontaneous emission of photons by the atoms. Since the QND interaction is a second order effect, spontaneous emission via photon emission of the excited state can be an eventual source of dephasing of the atomic states \cite{gao2022decoherence}. 
 %For the measurement operator, it scrambles the outcome such that only one measurement outcome is possible \cite{ilookeke2023}. It is made possible by shrinking the width of the measurement operator, thereby making the atoms' eigenvalues inaccessible except one which is immune to spontaneous emission. 
We note however that the MMES is in fact not the most sensitive state to dephasing by its nature \cite{byrnes2013fractality}.  Other types of entangled states such as Bell states composed of Schrodinger cat states are much more sensitive to dephasing and we expect such states are poor candidates for experimental realization.  On the other hand, the MMES as we consider here scales much better with the system size, and are a much more realistic prospect for experimental realization.  In a controlled experiment, where the detuning is large, effects arising from spontaneous emission can be controlled to be a small quantity.  
%Another potential challenge is to control the atom number fluctuations.  Experiments shot-to-shot will not have precisely the same atom numbers prepared in each trap, which may lead to additional errors in the readouts of particular quantities.  In terms of entanglement generation, extension of the theory to unequal atom numbers will simply result in entanglement that is generated corresponding to the smaller number of atoms between the two ensembles.  The remaining atoms are then not involved in the entanglement, and we expect that a MMES can still be formed. \textcolor{red}{We have found that the effect of the atom number fluctuation on the adaptive unitary is controllable in a sequential QND.}  
In summary, we consider the most critical threat to experimental realization of the MMES is the atomic dephasing that QND measurements induce.  However, this can be controlled, and with a careful choice of parameters, we believe dephasing effects can be minimized.

% We aim to study decoherence effects in the generation of the MMES and the application of these states as a resource state in the quantum information task of interest such as deterministic teleportation of atomic ensembles that is left as possible future work. 

%Hence our scheme is not affected by decoherence effects such as atom loss in the atomic ensembles. 

%Moreover, the QND scheme is robust against the photon loss i.e. photon loss has not a remarkable effect on the overall entanglement generation as noted in Ref. \cite{gao2022decoherence}. Another technical challenge could be to resolve the single photon counts in the experiments where it is used to project out the state as determined by the measurement operators (\ref{projzbasis}). We note that some progress towards this has been attained \cite{singlephoton}, nevertheless the convergence in our scheme does not affected by the photon counts as the adaptive unitary corrections are performed independently of the photon counts. 

%
%\begin{figure}[t]%
%	\includegraphics[width=\linewidth]{fig10b.eps}
%	\caption{Macroscopic maximally entangled state
%	}
%	\label{fig5}
%
%\end{figure}

\begin{acknowledgments}
This work is supported by the National Natural Science Foundation of China (62071301); NYU-ECNU Institute of Physics at NYU Shanghai; the Joint Physics Research Institute Challenge Grant; the Science and Technology Commission of Shanghai Municipality (19XD1423000,22ZR1444600); the NYU Shanghai Boost Fund; the China Foreign Experts Program (G2021013002L); the NYU Shanghai Major-Grants Seed Fund; Tamkeen under the NYU Abu Dhabi Research Institute grant CG008; Shanghai Frontiers Science Center of Artificial Intelligence and Deep Learning; and the SMEC Scientific Research Innovation Project (2023ZKZD55).
\end{acknowledgments}

\appendix

\section{Expression for transformation of Fock states through spin rotation}
\label{app:prob}
The Fock states $|k\rangle $ are eigenstates of the $ S^z $ spin operator, one can transform it to an arbitrary direction $| k \rangle^{(\theta, \phi)}$ as defined in Ref. \cite{timquantumoptics2020} where the matrix elements of the $S^y$ rotation are given by
\begin{align}
   &\langle k'|e^{-iS^y \theta/2}|k\rangle =  \sqrt{k!(N-k)!k'!(N-k')!}  \nonumber \\ &  \times    \sum_{n=\text{max}(k'-k,0)}^{\text{min}(k',N-k)}\frac{(-1)^n\cos^{k'-k+N-2n}(\theta/2)\sin^{2n+k-k'}(\theta/2)}{(k'-n)!(N-k-n)!n!(k-k'-n)!}
      \label{matrixelem}
\end{align}
where $|k\rangle=|k\rangle^{(z)}$.
\bibliography{ref}

\end{document}